\documentclass{article}

\usepackage{PRIMEarxiv}

\usepackage[utf8]{inputenc} 
\usepackage[T1]{fontenc}    
\usepackage{hyperref}       
\usepackage{url}            
\usepackage{booktabs}       
\usepackage{amsfonts}       
\usepackage{nicefrac}       
\usepackage{microtype}      
\usepackage{lipsum}
\usepackage{fancyhdr}       
\usepackage{graphicx}       
\graphicspath{{media/}}     

\usepackage{amsmath}
\usepackage[linesnumbered,ruled,vlined]{algorithm2e}
\usepackage{makecell}
\usepackage{multirow}
\usepackage{adjustbox}

\title{S\textsuperscript{2}M\textsuperscript{2}ECG: Spatio-temporal bi-directional State Space Model Enabled Multi-branch Mamba for ECG}

\author{
  Huaicheng Zhang \\
  School of Physics and Technology \\
  Wuhan University \\
  Wuhan, China\\
  \texttt{zhuaicheng@whu.edu.cn} \\
  \And
  Ruoxin Wang \\
  Pratt School of Engineering \\
  Duke University \\
  North Carolina, America\\
  \texttt{ruoxin.wang@duke.edu} \\
   \And
  Chenlian Zhou \\
  Faculty of Artificial Intelligence in Education \\
  Central China Normal University \\
  Wuhan, China\\
  \texttt{myphyllis@mails.ccnu.edu.cn} \\
    \And
  Jiguang Shi, Yue Ge \\
  School of Physics and Technology \\
  Wuhan University \\
  Wuhan, China\\
  \texttt{\{shijig, geeyue\}@whu.edu.cn} \\
  \And
  Zhoutong Li \\
  Huangpu Branch of Shanghai \\ Ninth People’s Hospital \\
  Shanghai, China\\
  \texttt{captain777@126.com} \\
  \And
  Sheng Chang, Hao Wang, Jin He, Qijun Huang\thanks{corresponding author} \\
  School of Physics and Technology \\
  Wuhan University \\
  Wuhan, China\\
  \texttt{\{changsheng, wanghao, jin.he, huangqj\}@whu.edu.cn} \\
}

\begin{document}
\maketitle

\begin{abstract}
As one of the most effective methods for cardiovascular disease (CVD) diagnosis, multi-lead Electrocardiogram (ECG) signals present a characteristic multi-sensor information fusion challenge that has been continuously researched in deep learning domains. Despite the numerous algorithms proposed based on different DL architectures, maintaining a balance among performance, computational complexity, and multi-source ECG feature fusion remains challenging. Recently, state space models (SSMs), particularly Mamba, have demonstrated remarkable effectiveness across various fields. Their inherent design for high-efficiency computation and linear complexity makes them particularly suitable for low-dimensional data such as ECG signals. This work proposes S\textsuperscript{2}M\textsuperscript{2}ECG, an SSM architecture featuring three-level fusion mechanisms: (1) Spatio-temporal bi-directional SSMs with segment tokenization for low-level signal fusion, (2) Intra-lead temporal information fusion with bi-directional scanning to enhance recognition accuracy in both forward and backward directions, (3) Cross-lead feature interaction modules for spatial information fusion. To fully leverage the ECG-specific multi-lead mechanisms inherent in ECG signals, a multi-branch design and lead fusion modules are incorporated, enabling individual analysis of each lead while ensuring seamless integration with others. Experimental results demonstrate that S\textsuperscript{2}M\textsuperscript{2}ECG achieves superior performance in the rhythmic, morphological, and clinical scenarios. Furthermore, its lightweight architecture ensures it has nearly the fewest parameters among existing models, making it highly suitable for efficient inference and convenient deployment. Collectively, S\textsuperscript{2}M\textsuperscript{2}ECG offers a promising alternative that strikes an excellent balance among performance, computational complexity, and ECG-specific characteristics, paving the way for high-performance, lightweight computations in CVD diagnosis.
\end{abstract}


\section{Introduction}
Cardiovascular diseases (CVDs) represent a significant global health burden, affecting populations worldwide and posing substantial threats to public health. As a non-invasive, cost-effective diagnostic modality, the electrocardiogram (ECG) has become an indispensable clinical tool for rapid assessment of cardiac function. By capturing and analyzing the heart's electrical activity, this technology enables detection of key pathophysiological features associated with various CVDs, as demonstrated in recent studies \cite{hannun2019cardiologist, sangha2022automated}. Notwithstanding these advantages, manual interpretation of ECGs remains time-consuming and prone to human error. To address these limitations, advanced deep learning (DL) algorithms have been increasingly implemented to automate the diagnostic process, offering the potential for improved efficiency and accuracy in clinical decision-making. \par
Diverse deep learning (DL) paradigms have demonstrated transformative impacts across multiple domains, including computer vision (CV), natural language processing (NLP), intelligent healthcare, etc. These advancements underscore the domain-specific adaptability of DL architectures, where certain models exhibit distinct advantages in particular application scenarios. Different deep learning architectures are tailored to specific tasks. In NLP, the Transformer architecture excels due to its self-attention mechanism\cite{vaswani2017attention}, which effectively captures long-range dependencies in sequential data. For CV, convolutional neural networks (CNNs) dominate because their local receptive fields and translation invariance align perfectly with the hierarchical spatial patterns in images. In the context of ECG analysis, a wide range of architectural variations have been explored, including CNNs, transformers, recurrent neural networks (RNNs), and graph neural networks (GNNs), etc. \par
However, existing frameworks may not be optimal, as they often fail to simultaneously model long-term dependencies, capture spatio-temporal features, and maintain the computational efficiency required for cardiac signal analysis. In brief, CNNs struggle with long-range temporal dependencies due to their localized filters, Transformers face quadratic complexity bottlenecks in long sequences. RNNs suffer from vanishing gradients and slow sequential processing, while GNNs require predefined graph structures that may not accurately describe physiological relationships. More critically, the high computational complexity of some state-of-the-art architectures creates significant barriers to real-time deployment on resource-constrained edge devices, which are important for ECG daily monitoring. These limitations collectively underscore the need for novel paradigms that simultaneously address algorithmic efficacy and computational efficiency in cardiac signal analysis.

Recently, an innovative architecture naming Mamba is proposed, which is based on structured state space sequence models (SSMs). SSMs offer a compelling alternative: they combine linear-time complexity for long sequences, selective state transitions for adaptive feature focus, and continuous-time modeling that inherently handles irregular sampling - effectively addressing the limitations of previous approaches. It not only simplifies the neural network architectures without attention mechanisms but also introduces long-range recognition domains to play similar roles of such attention. In addition, convolutions are made more efficient\cite{gu2023mamba}. Incorporating the features of CNNs and RNNs, Mamba is regarded as a hardware-aware parallel algorithm in recurrent mode with linear computational complexity. Since RNNs' sequence manner is changed to a parallel one, computational efficiency, and long-range information could be acquired simultaneously. It could be said that Mamba is naturally appropriate for time-series data. Therefore, these properties make Mamba particularly promising for ECG analysis. It demonstrates exceptional potential for cardiac signal processing: its input-dependent state transitions naturally capture arrhythmia morphology, while the linear scaling enables efficient analysis of long-term recordings. \par 

Motivated by these advantages, a novel Mamba-based framework tailored for intelligent ECG diagnosis is introduced. The proposed architecture incorporates multi-branch bi-directional state space processing to address the unique challenges of ECG interpretation. The key contributions of this work are as follows:
\begin{itemize}
	\item Domain-Specific Architecture Design: A novel cardiovascular disease (CVD) diagnosis framework that integrates ECG multi-branch mechanisms with spatio-temporal bi-directional SSMs is presented. This architecture leverages Mamba's linear complexity and long-range modeling capabilities while accounting for the multi-lead nature of clinical ECG recordings.
	\item Efficient Long-Range Dependency Modeling: By adopting structured state space representations, the proposed model inherits Mamba's ability to capture long-distance temporal dependencies without the quadratic complexity of self-attention. The bi-directional processing paradigm further enhances rhythmic pattern recognition, critical for detecting arrhythmias and other time-dependent CVD manifestations.
	\item Hierarchical Feature Extraction: Through ECG segment tokenization and bi-directional scanning, the model achieves local information aggregation analogous to convolutional kernels while maintaining global context awareness. This hybrid approach enables simultaneous morphological analysis and rhythm interpretation, addressing two core dimensions of ECG diagnosis.
	\item Multi-Lead Information Fusion: The architecture explicitly models the spatial relationships between ECG leads through a multi-branch design, where each lead-specific Mamba encoder extracts unique features. Subsequent fusion modules integrate these multi-dimensional representations to capture lead-interdependent pathophysiological patterns.
	\item Clinically Deployable Solution: Comparing to existing methods on performance and light-weight aspects, the proposed S\textsuperscript{2}M\textsuperscript{2}ECG achieves superior and favorable performance in rhythmic, morphological, and clinical scenarios. Moreover, S\textsuperscript{2}M\textsuperscript{2}ECG nearly has one order of magnitude fewer parameters than most existing methods. So that it can be deployed more easily in clinical practices.
\end{itemize}

\section{Related Work}
\subsection{Deep Learning in ECG}
With the development of deep learning technologies, there emerge diverse intelligent diagnosis methods based on deep learning. Specifically, convolutional neural networks (CNNs) have established themselves as one of the most conventional and classic architectures in medical image analysis. Their hierarchical feature extraction capability aligns well with the multi-scale characteristics of biomedical signals. Usually, residual designations and 1-dimensional convolution are employed\cite{he2016deep, hannun2019cardiologist, avetisyan2024deep}. Since their advantages in processing time-series data, RNNs are prevalent around 2010s, especially long-short term memory (LSTM)\cite{greff2016lstm}. There exist many hybrid models based on CNNs and RNNs\cite{yao2020multi, he2019automatic, chen2022automated, xie2022multilabel}, they integrates CNN’s local receptive fields and RNN’s sequential modeling for ECG. \par 
While as the proposition of Transformers, self-attention mechanisms are proven more efficient than RNNs. Thereby RNNs are replaced by Transformers in many researches\cite{vaswani2017attention, ji2024msgformer}. Except for such universal models, some ECG-specific models focus on the unique features of ECGs. For instance, the multi-branch mechanisms introduce lead-wise designation for Myocardial Infarction(MI) detection\cite{liu2019mfb}, graph models introduce topological relationships among different leads\cite{zhang2023st, zhang2024honest}, ECG-specific attention mechanisms introduce particular recognition as it is incorporated in 1-D CNN\cite{guhdar2025advanced}, etc. Nevertheless, as indicated in prior discussions, the pursuit of diagnostic accuracy often comes at the cost of computational complexity. A model with better balance between complexity and performance is worth researching.

\subsection{State Space Sequence Models}
As a traditional mathematical theory, SSMs can capture temporal data dependencies effectively by hidden states. Recent theoretical breakthroughs in structured state-space duality (SSD) endow SSMs with enhanced capabilities to model both local patterns and global context\cite{qu2024survey}. By establishing a rigorous mathematical framework connecting structured SSMs and attention mechanisms, Mamba bridges the gap between recurrent and Transformer-based paradigms. This duality allows selective retention of critical signal features while maintaining linear computational complexity—a key advantage for processing long ECG recordings\cite{gu2023mamba}. \par
Therefore, Mamba-based models are widely utilized in different domains, e.g., language, vision, time series. As representatives, MambaByte leverages the advantages of Mamba in capturing long-range dependencies for token-free language models\cite{qu2024survey, wang2024mambabyte}. Vision adaptations like Vim and VMamba reveal architectural flexibility: while Vim processes flattened image patches as 1D sequences \cite{zhu2024vision}, VMamba preserves 2D spatial inductive bias through directional scanning \cite{liu2024vmamba}—a strategy potentially relevant to ECG's quasi-2D lead arrangements. Except for them, in vision domains, Pan-Mamba\cite{he2025pan} and Polyp-Mamba\cite{zhu2025polyp} are proposed as well. Pan-Mamba  leverages the efficiency of the Mamba model in global information modeling\cite{he2025pan} and Polyp-Mamba incorporates Mamba and ResNet for precise polyp segmentation\cite{zhu2025polyp}. They In semi-supervised learning, Semi-Mamba-UNet is proposed. It integrates a visual Mamba-based U-shaped encoder–decoder architecture with a conventional CNN-based UNet.\cite{ma2024semi} In the domains of image retrieval, Mamba is also incorporated in Transformer. FMTH, for example, constructs a framework for frequency decoupling enhancement and Mamba depth extraction-based feature fusion in Transformer hashing image retrieval\cite{chen2025frequency}. TimeMachine harnesses Mamba to capture enduring patterns in multivariate time-series data\cite{qu2024survey,ahamed2024timemachine}. For time series as well, TSCMamba integrates diverse features, including spectral, temporal, local, and global features, to obtain rich, complementary contexts for time-series classification with Mamba\cite{ahamed2025tscmamba}. In addition, in the domains of traffic forecasting, MGCN is proposed. It's a Mamba-integrated spatio-temporal graph convolutional network for long-term traffic forecasting\cite{lin2025mgcn}. Similarly, ST-MambaSync, integrates Mamba and Transformer technologies with balanced computational cost and higher accuracy for traffic prediction\cite{shao2025st}.\par 
However, ECG's unique multi-lead topology poses distinct challenges. Unlike vision data with explicit 2D grids or uni-variate time series, 12-lead ECGs exhibit implicit spatio-temporal correlations across leads (e.g., augmented limb vs. chest leads) and time-frequency characteristics (e.g., P-QRS complex (QRS)-T waveform morphology). The research of promoting Mamba-based models in ECG domains is still under-explored. In other words, unique features of ECGs like multi-lead mechanisms, spatio-temporal relationships, time-frequency characteristics could be included in the improvements.

\subsection{SSMs in CVD Diagnosis}
Considering the success of Mamba-based models in other domains, some research focusing on applied Mamba to CVD diagnosis has being emerging. The recent proliferation of Mamba-based architectures in cardiovascular disease (CVD) diagnosis highlights growing recognition of their temporal modeling capabilities. ECGMamba employs a bidirectional SSM to CVD diagnosis. ECGMamba pioneers bidirectional state-space modeling for multi-beat ECG analysis, claiming 15\% faster inference than Transformer counterparts while maintaining competitive accuracy on the PhysioNet Challenge 2021 dataset\cite{qiang2024ecgmamba}. However, its channel-agnostic processing of 12-lead data may overlook the diagnostic significance of lead-specific waveform variations (e.g., ST-segment elevation in precordial leads). Different from ECGMamba's multi-beat level, MambaCapsule focuses on single-beat level data. In that work, Mamba block and Capsule networks are employed for feature extraction and prediction\cite{xu2024mambacapsule}. While localized anomaly detection is supported (e.g., premature ventricular contractions), its reliance on precise R-peak alignment limits applicability to noisy ambulatory recordings. Except for CVD diagnosis, Mamba models are applied to signal denoising, such as MSECG\cite{lin2024msecg} and MECG-E\cite{hung2024mecg}. Generally, multi-beat level ECG data is more closed to clinical practices, which is without R-peak detection, heartbeat segmentation, etc. \par

\begin{table*}[htbp]
	\centering
	\caption{Comparison of related models tailored for CVD diagnosis.}
        \adjustbox{width=\textwidth}{
	\begin{tabular}{ccccccccc}
		\toprule
		Method & \multicolumn{1}{l}{Year} & Architecture & Task  & Data  & \makecell{Multi\\Branch} & \makecell{Paralle\\-lization}  & \makecell{Long-range\\Dependencies} & Complexity \\
		\midrule
		\cite{hannun2019cardiologist} & 2019  & CNN   & Classifying & 1-lead &       & $\checkmark$ &       & linear \\
		\cite{he2019automatic} & 2019  & CNN+RNN & Classifying & 12-lead &       &       & $\checkmark$ & linear \\
		MFB-CBRNN\cite{liu2019mfb} & 2019  & CNN+RNN & MI detection & 12-lead & $\checkmark$ &       & $\checkmark$ & linear \\
		\cite{yao2020multi} & 2020  & CNN+RNN & Classifying & 12-lead &       &       & $\checkmark$ & linear \\
		\cite{chen2022automated} & 2022  & CNN+RNN & Classifying & 12-lead &       &       & $\checkmark$ & linear \\
		\cite{xie2022multilabel} & 2022  & CNN+RNN & Classifying & 12-lead & $\checkmark$ &       & $\checkmark$ & linear \\
		ST-ReGE\cite{zhang2023st} & 2023  & GNN   & Classifying & 12-lead &       & $\checkmark$ &       & linear \\
		\cite{avetisyan2024deep} & 2024  & CNN   & Classifying & 12-lead &       & $\checkmark$ &       & linear \\
		MSGformer\cite{ji2024msgformer} & 2024  & CNN+Trans. & Classifying & 12-lead &       & $\checkmark$ & $\checkmark$ & quadratic  \\
		ECGMamba\cite{qiang2024ecgmamba} & 2024  & SSM   & Classifying & 12-lead &       & $\checkmark$ & $\checkmark$ & linear \\
		MambaCapsule\cite{xu2024mambacapsule} & 2024  & SSM   & Classifying & 1-lead &       & $\checkmark$ & $\checkmark$ & linear \\
		MSECG\cite{lin2024msecg} & 2024  & SSM   & Filtering & 12-lead &       & $\checkmark$ & $\checkmark$ & linear \\
		MECG-E\cite{hung2024mecg} & 2024  & SSM   & Filtering & 12-lead &       & $\checkmark$ & $\checkmark$ & linear \\
		\cite{guhdar2025advanced} & 2025  & CNN   & Classifying & 1-lead &       & $\checkmark$ &       & linear \\
		S\textsuperscript{2}M\textsuperscript{2}ECG & 2025  & SSM   & Classifying & 12-lead & $\checkmark$ & $\checkmark$ & $\checkmark$ & linear \\
		\bottomrule
	\end{tabular}%
	\label{tab:addlabel2}%
    }
\end{table*}%

These advancements, however, primarily adapt existing Mamba paradigms without addressing ECG's unique multi-lead spatio-temporal inter-dependencies. Therefore, it's worth researching on an ECG-specific Mamba model, which is proposed in this work. For intuitively analyzing various properties, related models tailored for ECG are listed in Table 1. These methods are with different architectures and might be utilized for various tasks. The data utilized is usually 1-lead or 12-lead. Despite these differences, the evaluation of models is more related to the model properties. There are several properties for comparisons. Firstly, the multi-branch designation. It's a unique designation for multi-lead ECG signals that has been applied for some ECG-specific models\cite{liu2019mfb, yao2020multi}. Such structure provides a dual process to analyze each lead individually and fuse information from different leads. It is validated that such structure can effectively integrate spatio-temporal information and demonstrates good performance. With respect to the parallelization, RNN-based models merely support serial calculations, it results in long time for training and inference. In addition, while CNNs demonstrate excellent local receptive field properties, they suffer from limited long-range dependency capture. Consequently, they are inferior in temporal analyzing which is significantly dependent on long-range dependency. Finally, transformer-based models are with good parallelization and long-range dependencies due to their self-attention mechanisms. But such mechanisms bring quadratic computational complexity, resulting in high memory utilized in long sequence scenarios. Therefore, there is a research gap for an ECG-specific high-efficiency method with satisfactory performances and low computational complexity. To bridge this gap, S\textsuperscript{2}M\textsuperscript{2}ECG is introduced as a novel solution. \par

\begin{figure*}[htb]
	\centering	
	\includegraphics[width=\textwidth]{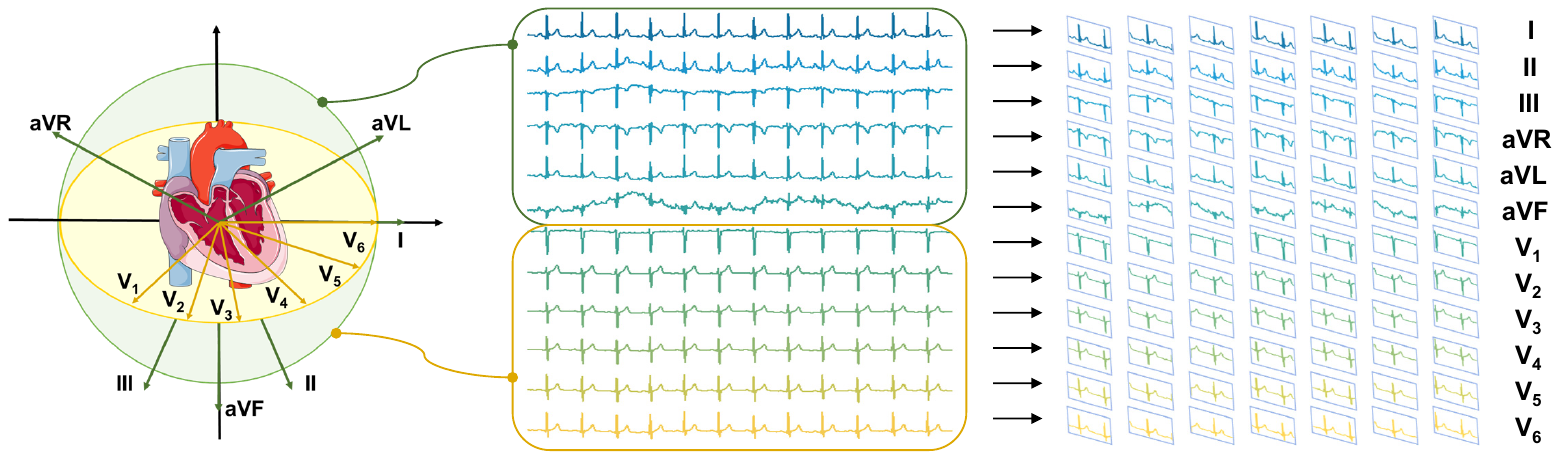}
	\caption{The illustration of lead systems and process of ECG signals slicing.}\label{Figure 1}
\end{figure*}

\section{Methodology}
\subsection{Preliminaries}
The SSM is a traditional mathematical framework utilized to depict the dynamic behavior of a system over time\cite{qu2024survey}, thus it's utilized in continuous scenarios, as shown in Eq. (1), where $h'(t)$ is the derivative of current state $h(t)$, $x(t)$ and $y(t)$ are the inputs and outputs. $A\in\mathbb{R}^{N\times N}$ is the state transition matrix, $B\in\mathbb{R}^{N\times 1}$ and $C\in\mathbb{R}^{1\times N}$ are the input and output matrix. $D\in\mathbb{R}$ is the command coefficient describing how the outputs are directly affected by the inputs. 

\begin{flalign}
	&&
	h'(t)=Ah(t)+Bx(t)\qquad y(t)=Ch(t)+Dx(t)
	&&
\end{flalign}

However, for data with aggregated features like ECG, language, and pictures, continuous SSMs should be discretized and made usable in discrete scenarios. For this aim, Zero-Order Hold (ZOH) is employed, then the formulas can be changed to Eq.(2), where $\bar{A}=e^{\bigtriangleup A}$, $\bar{B}=(\bigtriangleup A^{-1})(e^{\bigtriangleup A}-I)$, and $\bigtriangleup=[t_{k-1},t_k]$. After discretization, computational efficiency is reinforced since the computations are simplified.

\begin{flalign}
	&&
	h_k=\bar{A}h_{k-1}+\bar{B}x_k\qquad y_k=Ch_k
	&&
\end{flalign}

After that, a global convolution is introduced to aggregate the features, yielding the outputs, as described in Eq. (3), where $\bar{K}\in \mathbb{R}^M$ denotes a convolution kernel and $x$ is the input.

\begin{flalign}
	&&
	\left\{
	\begin{aligned}
		\bar{K}=(C\bar{B},C\bar{AB},...,C\bar{A}^{M-1}\bar{B}) \\
		y=x*\bar{K}
	\end{aligned}
	\right.
	&&
\end{flalign}

\begin{figure*}[htbp]
	\centering	
	\includegraphics[width=\textwidth]{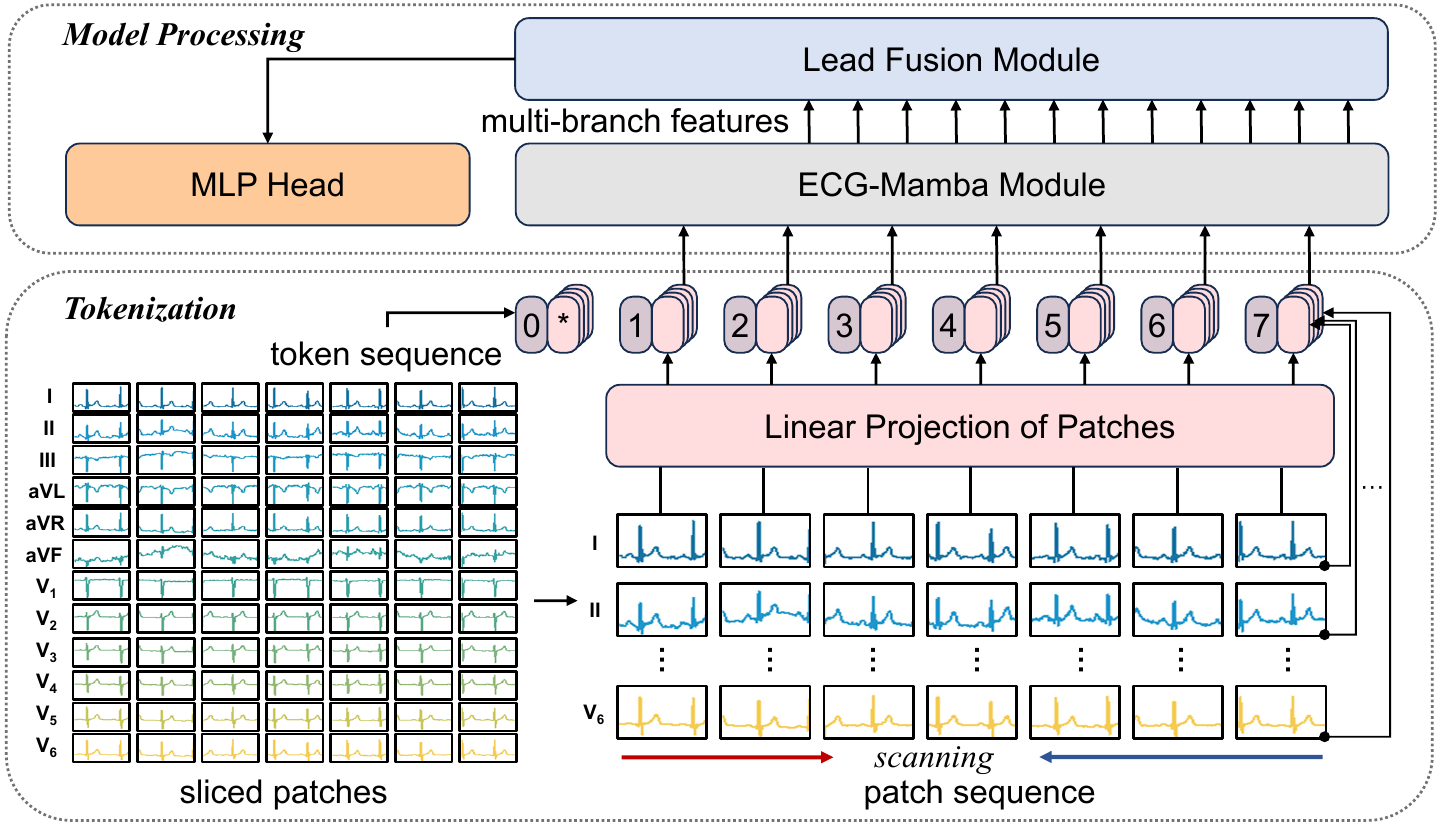}
	\caption{The workflow of S\textsuperscript{2}M\textsuperscript{2}ECG, including segment tokenization and model processing.}\label{Figure 2}
\end{figure*}

\subsection{Segment Tokenization}
In Mamba-based models, the continuous SSMs have been modified to a discrete style. The ECG signals are continuous, thereby they should also be tokenized into discrete tokens. After segment tokenization, ECG becomes a discrete sequence suitable for Mamba block processing. It enhances Mamba's spatial awareness of ECG signals, thus the ability of spatial modeling will be enhanced. The reason for not utilizing continuous SSMs to analyze continuous ECGs is that ECGs not only have temporal trends but also spatial features. In other words, local information aggregation is needed, so that local information can be analyzed as a union. Segment tokenization transforms the analysis level from a time step to a local time-series segment. \par
Standard ECG data is composed of 12 synchronous signals on 12 leads, sourcing from the limb and chest systems, as illustrated in Fig. (1). Although each ECG signal is one-dimensional data, there are other 11 synchronous signals on other leads. In other words, the lead level can be regarded as the channel level. Thereby ECG data can be regarded as 2-dimensional data with 12 channels, and the height in each channel is 1. Such data could be called sub-2-dimensional data. The lead systems are important because they point to the activities of particular heart regions. Only combining different leads can yield a comprehensive analysis of an ECG. Different leads contain diverse information, thus it's unsuitable to directly aggregate the information of the whole ECG. A feasible approach is analyzing signals on each lead individually, and aggregating them finally. Therefore, in the segment tokenization stage, signals on different leads should be sliced individually and synchronously.

Inspired by vision Transformers (ViT), ECGs are segmented into time-series pieces, and each piece is represented as a token. Note that the segment position in each lead is the same, considering that each ECG has 12 leads. The segment tokenization process is illustrated in Fig. (2). Considering that each lead's token sequence is analyzed individually, individual classification tokens (CLS) should be added at the start and the end of each token sequence. Finally, the token sequences could be formulated as Eq. (4), where $T_i$ denotes the token sequence of lead $i$, N denotes the number of tokens per sequence, $t_{CLS}$ denotes the CLS token, $d$ denotes the scanning direction which could be forward and reverse, and $E_{pos}$ denotes the position embedding of the token sequence. 
\begin{flalign}
	&&
	T_{d,i}=[t_{CLS},t_p^{1,i},t_p^{2,i},...,t_p^{N,i}]+E_{pos}\quad(i=1,2,...,12)
	&&
\end{flalign}

\subsection{ECG-specific Multi-branch Architecture}
After segment tokenization, the ECG signal in each lead is segmented into several tokens, the tokens are the same length and the number of tokens for each lead is the same. Since multi-branch characteristics of ECGs are introduced to S\textsuperscript{2}M\textsuperscript{2}ECG, the tokens of each lead are sent to individual ECG-Mamba encoders. Inspired by the success of multi-branch designation in CNN-based CVD diagnosis models\cite{liu2019mfb}, the same mechanism is employed in S\textsuperscript{2}M\textsuperscript{2}ECG. Thus, 12 different classification tokens are needed for 12 sequences, which are represented as * in Fig. (2). Individual encoders would yield lead-specific parameters, which are helpful extract features of particular leads. \par 
Since the output of each individual encoder merely includes lead-specific information, only temporal characteristics are extracted in this stage. However, intra-lead information also matters because it reflects the interaction of different heart regions, which are beneficial in lesion localization and spatial-related disease diagnosis. To synthesize intra-lead information, the output of ECG-Mamba modules is sent into a lead-fusion module. Though several convolution layers with 1 kernel and 12 channels, synchronous information is shared during corresponding tokens. Such designation only introduces spatial information sharing and avoids further aggregation of temporal features, not resulting in the forgetting of existing knowledge. \par 

\subsection{Bi-directional Scanning}
Another designation is that the scanning of token sequences is bi-directional. The scanning mechanism is the core design for its processing of multi-dimensional data. Its main function lies in serializing high-dimensional data and capturing global information while maintaining linear computational complexity. The scanning direction can determine the order of information capture and thereby construct global dependencies. In this work, bi-directional scanning is employed. From the aspect of algorithm designation, forward scanning processes ECG segments along the original order, capturing local features and initial context. Reverse scanning processes the sequence in reverse order, complementing the portion of global information omitted in forward scanning. Finally, integrating the forward and reverse scans enhances spatial continuity. From the medical aspects, based on existing experience, later time-series segments could also provide extra knowledge to help the analysis of former segments. This is also why bi-directional LSTM is prevalent in CVD diagnosis several years earlier. In CVD diagnosis, not all CVDs can be diagnosed by a single heartbeat. In most cases, the diagnosis of CVD requires relying on the long-term recording of heartbeats. Many diseases will exhibit variability sometime after a certain heartbeat, which is the time delay and potentiality of CVD. This is also the reason why the professional ECG machine (Holter monitor) needs to be worn for a long time in order to better record the connections between heartbeats\cite{gibson2007diagnostic}. Thus, bidirectional designations are also employed in the Mamba-based models. Specifically, the token sequences are sent to ECG-Mamba encoders twice, once forward and once reverse. Finally, these two outputs are concatenated as the final outputs of ECG-Mamba modules.

\begin{figure*}[htbp]
	\centering	
	\includegraphics[width=\textwidth]{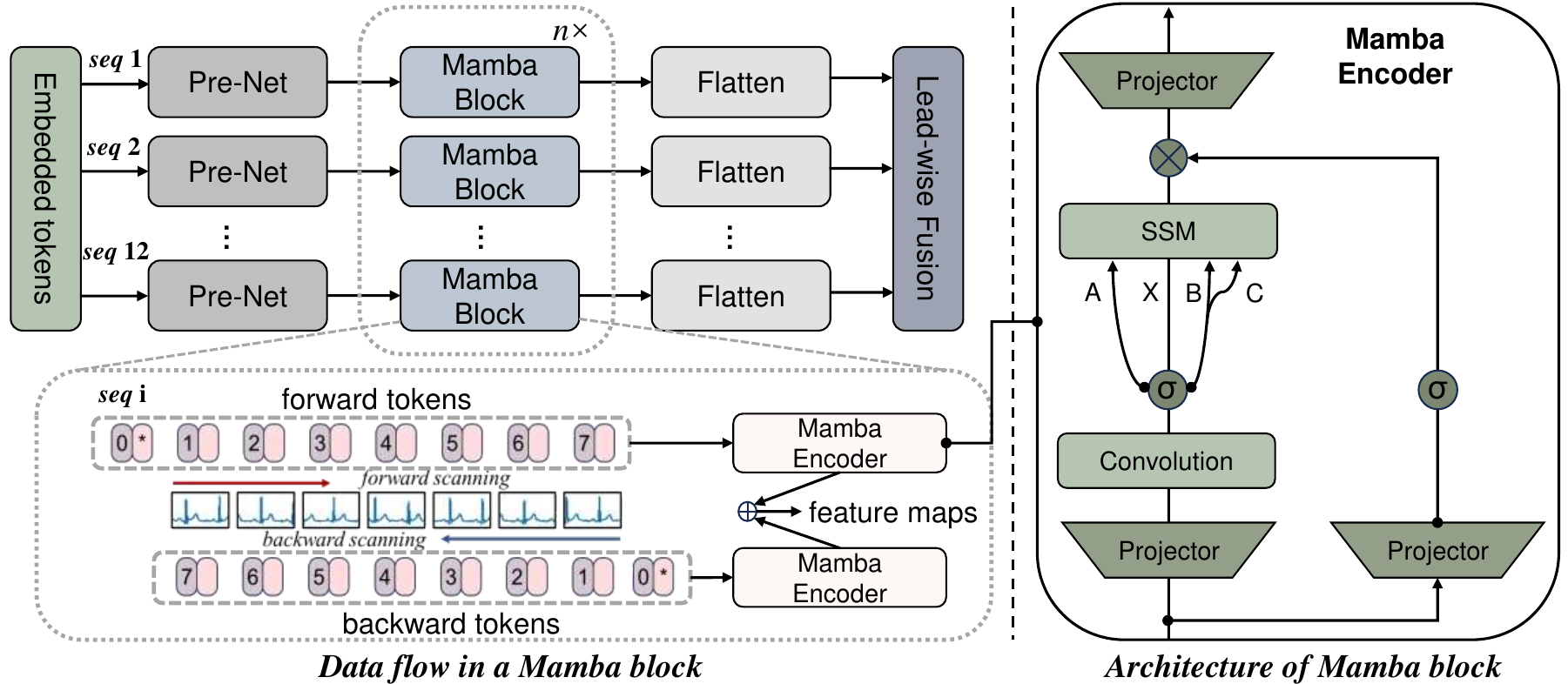}
	\caption{The architecture and data workflow of ECG-Mamba module in S\textsuperscript{2}M\textsuperscript{2}ECG.}\label{Figure 3}
\end{figure*}

The proposed segment tokenization and scanning mechanisms exhibit functional analogies to convolutional kernels while introducing key advancements. Similar to convolution's local receptive fields, segment tokenization partitions signal into non-overlapping patches to extract localized features through linear projection. Bi-directional scanning mechanisms further mimic large-kernel convolutions by aggregating global spatial relationships, yet fundamentally differ in two aspects: 1) Dynamic parameterization – Unlike static convolutional weights, SSM matrices adaptively adjust based on input content, prioritizing salient regions like ECG wave peaks; 2) Hardware-aware efficiency – the bi-directional scanning strategy achieves full-signal receptive fields with linear complexity, outperforming multi-layer CNN stacks in both speed and memory efficiency. This hybrid design effectively bridges the strengths of inductive bias-free sequence modeling and spatially-aware feature learning, particularly advantageous for ECG data.\par

The architecture of S\textsuperscript{2}M\textsuperscript{2}ECG is illustrated in the Fig. (3). It could be seen that the token sequences are first analyzed by a Pre-Net, which projects the input tokens into feature spaces with particular dimensions. Then they are analyzed by bi-directional Mamba blocks and transformed into distilled features. After flattening these features and recording them in corresponding channels (leads), the final features could be sent to the lead-fusion modules. Thus, in the process from token sequences to Mamba outputs, each lead is individual, and lead-specific features could be fully extracted. \par 

\subsection{Lead Fusion Module and Classifying Head}
After being processed by Mamba blocks, the output features are sent to the lead fusion module. Different from simple concatenation, this module contains a feed-forward network (FFN)\cite{vaswani2017attention} to enrich the temporal feature diversity and a squeeze-and-excitation network (SENet)\cite{hu2018squeeze} to provide spatial channel-wise attention. Finally, the processed features are sent to the classifying head for categorizing, which is composed of multi-layer perceptrons. The details of the lead fusion module and the Classifying head are illustrated in Fig. (4), where B, dim, and class denote the batch size, the feature dimensions, and the number of classes respectively.\par
Specifically, the FFN expands and recovers the feature dimensions on a temporal scale. The hidden dimensions of the FFN are twice the input dimension. It enriches the temporal feature space and complements the nonlinear transformation of the time scale. After that, to acquire spatial attention, SENet is introduced to weigh different channels (leads). The features are squeezed on the temporal scale first and then down-scaled on the lead scale, i.e., the spatial scale. The down-scaling rate is 12, in other words, inputs originating from 12 leads are down-scaled to 1 channel. Through excitation, the features are recovered and the channel-wise attention is yielded, which is scaled to the input for weighting. Such SENet designation provides channel-wise attention to determine the significance of various leads, it results in a channel-wise spatial feature fusion. Finally, the lead fusion module fuses the 12 inputs from Mamba blocks in a spatio-temporal manner. \par
The output features are sent to the classifying head for categorizing. The classifying head is composed of linear layers for gradual mapping. The features are mapped to 64 dimensions first, and after activation and batch normalization, they are concatenated on a temporal scale. Finally, the concatenated features are classified into output classes. \par

\begin{figure*}[htbp]
	\centering	
	\includegraphics[width=0.8\textwidth]{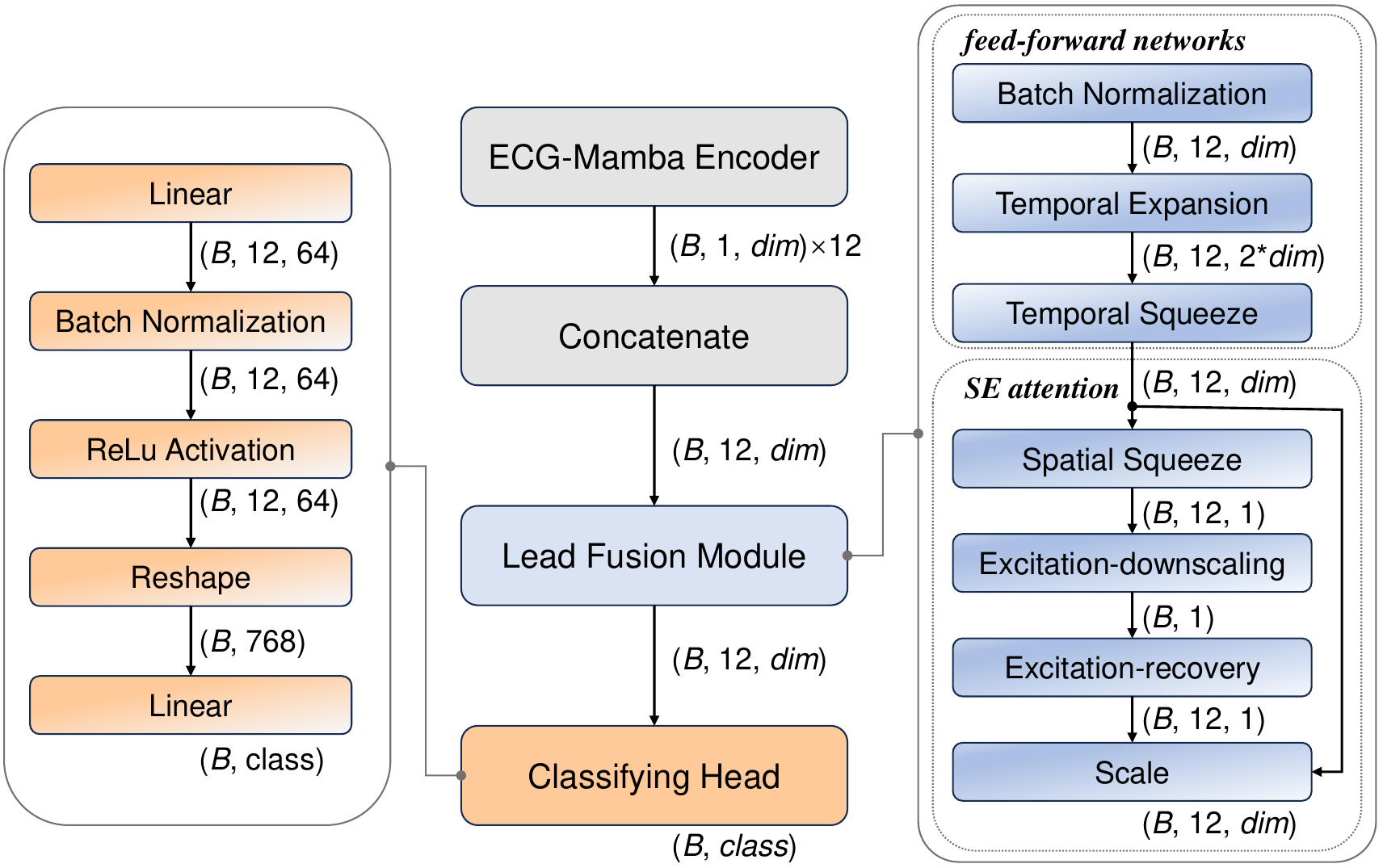}
	\caption{The lead fusion module and classifying head of S\textsuperscript{2}M\textsuperscript{2}ECG.}\label{Figure 4}
\end{figure*}

\begin{algorithm}[htbp] 
	\caption{The workflow of S\textsuperscript{2}M\textsuperscript{2}ECG}
	\begin{small} 
		\BlankLine 
		\KwIn{token sequence of all leads $T_{i,l}\quad(i=1,2,...,12)$, linear classifier $CLS$, sequence length $L$, SSM block $SSM$, parameters $\bigtriangleup_0$ and $\bigtriangleup_A$.} 
		\KwOut{output logits $l_{out}$}
		\For{$l\in range(L)$}{
		\For{$i\in [1,2,...,12]$}{
			$T_{i,l}'=NORM(T_{i,l})$\;
			$/*d\leftarrow direction, i\leftarrow lead\ index, $\
			$l \leftarrow sequence\ index*/$\;
			$x_{i}=Linear(T_{i,l}')$\;
			$z_{i}=Linear(T_{i,l}')$\;
			\For{$d \in [forward,backward]$}{
				$/* Apply\ directions */$\;
				$x_{d,i}=x_i$\;
				$z_{d,i}=z_i$\;
				$x_{d,i}'=SiLu(Conv1d(x_{d,i}))$\;
				$B_{d,i}=Linear(x_{d,i})\qquad C_{d,i}=Linear(x_{d,i})$\;
				$/* Activation */$\;
				$\bigtriangleup=log(1+exp(Linear(x_{d,i})+\bigtriangleup_0))$\;
				$/* Discritinization */$\;
				$\bar{A_{d,i}}=discrete(\bigtriangleup,\bigtriangleup_A)$\;
				$\bar{B_{d,i}}=discrete(\bigtriangleup,B_{d,i})$\;
				$y_{d,i}=SSM(\bar{A_{d,i}},\bar{B_{d,i}},C_{d,i})(x_{d,i}')$\;
				$y_{d,i}'=y_{d,i}\cdot{SiLu(z_{d,i})}$
			}
			$y'=y_{forward,i}'+y_{backward,i}'$\;
			$T_{i,l+1}=Linear(y')+T_{i,l}$
		}
		$T_{i}=\sum_{1}^{L}T_{i,l}$
		}
		$T=Lead\ Fusion\ Module(T_{i})\qquad(i=1,2,...,12)$\;
		$T=Linear(T)$\;
		$l_{out}=CLS(T)$\; 
		{
			return $l_{out}$
		}
	\end{small} 
\end{algorithm}

The pseudo-code of the whole process is shown in Algorithm (1). For the ECG token sequence of each lead, a normalization is introduced first to eliminate internal covariate shifts. Then, the processed sequences are sent to linear layers and projected into the input of the state space $x$ and the gating signal $z$. After that, the bi-directional scanning is initiated, and the inputs are scanned in forward and backward directions, respectively. For each direction, the inputs are projected to continuous time matrix $B$ and output projection matrix $C$. At the same time, time step adaptive parameters $\bigtriangleup$ are prepared. With $\bigtriangleup$, the state transfer matrix $A$ and the inputs projection matrix $B$ can be discretized, and the continuous time system is converted to the discrete-time system. Thereby these parameters could be sent to the SSM modules. In this step, transformed features are generated. The outputs of SSMs in both directions are then concatenated, and the processed outputs from 12 leads are sent to the lead fusion module to fuse features temporally and spatially. Finally, the fused features are utilized for classification. Except for the designation of model architectures, the detailed parameters would significantly affect the performance. For example, the length of tokens, the number of Mamba blocks, the dimension of feature spaces, the step length in scanning, etc. For different kinds of CVDs, the best hyper-parameters might vary. This issue will be analyzed in the experiment part. \par

\section{Experiments and Discussion}
\subsection{Materials}
In this research, four datasets are introduced to simulate different scenarios. Specifically, considering the CVD categories and data origins, PTB-XL\cite{wagner2020ptb}, Chapman\cite{zheng202012}, and the Shanghai Ninth Peoples’ Hospital (SNPH) datasets are introduced to simulate morphological CVD diagnosis, rhythmic CVD diagnosis, and clinical practices, respectively. Additionally, the
Ningbo First Hospital (NFH) dataset\cite{zheng2020optimal} is introduced to evaluate S\textsuperscript{2}M\textsuperscript{2}ECG's generalization capabilities in clinical practices. The details and data partitioning are shown in Table 2. CVDs in PTB-XL like MI and STTC can easily diagnosed by morphological features, while those in Chapman correspond to rhythmic ones. All four datasets were utilized for training, validation, and testing. The model was trained separately on each dataset, validated accordingly, and ultimately evaluated on their respective test sets. Since generalization experiments could only be conducted between databases with the same categories, Chapman and NFH databases are utilized for verify model performances across databases. In these situations, the model can be trained on one database and test on another one.\par 
Clinical databases are crucial to evaluate the model deployability in clinical practices. Thus, SNPH and NFH databases are introduced in this work. Concerning SNPH, ECGs in that private database are collected from clinical ECG machines and desensitized, removing the patient’s identification information. Note that the approval of the SNPH database is exempted according to the “Ethical Review of Life Science and Medical Research Involving Human Beings” issued by the National Health Commission of the People’s Republic of China\cite{national2023circular} and "Notice on further improving the ethical review of life sciences and medical research involving human beings" issued by Shanghai Health Care Commission\cite{shanghai2023notice}: "Second, use of anonymized information data for research is exempt from ethical review"\cite{zhang2023sigxcl}. The native and raw data are collected from clinical ECG machines of the Huangpu Branch of Shanghai Ninth People's Hospital. All the data is desensitized, i.e., the patient’s identification information has been removed. Only ECG data and the diagnosis are preserved. After these processes, the ECGs are in the same form as those in open-access databases. Finally, proposed models can be evaluated more comprehensively with different categories of CVDs and scenarios. \par
With respect to the details of the SNPH database, it is a private multi-label database that contains 51523 labeled ECGs and 23588 unlabeled ones originated from 75111 patients. Within the 51523 patients, the number of female patients was 26482, male patients were 16298, and the gender of the remaining 8559 was unknown. And the age distribution is: 25.62\% under 50 years, 13.33\% 50-60 years, 24.36\% 60-70 years, and 19.91\% over 70 years. The ECGs in the dataset were acquired via the Nalong aECG-18U recorder\cite{nlhealth2025nalong}. The data annotation was derived from clinical diagnosis, performed by expert doctors from the Huangpu Branch of Shanghai Ninth People's Hospital. The aECG-18U electrocardiograph automatically generates a preliminary detection report, which is then verified and corrected by doctors to ensure annotation accuracy. The original ECG data has a sampling rate of 500 Hz. Since the research scope of this work is single-label classification, the multi-label data from the SNPH database is excluded. It means that only single-label data is retained. And the unlabeled data is excluded as well. Then, 5 main CVDs are selected as the research objects, as shown in Table 2. Finally, there are 9491 recordings are utilized for research. Since these data originate from clinical diagnostics and the selected categories represent predominant classifications within the single-label ECGs, they demonstrate strong representatives of practical clinical scenarios. \par
As for the NFH database, it's a large and expertly labeled open-access clinical dataset consisting of 12-lead ECGs with a sampling rate of 500 Hz from 40,258 patients\cite{zheng2020optimal}. Its age groups with the highest prevalence were 51–60, 61–70, and 71–80 years representing 19.8\%, 24\%, and 17.3\% respectively. Its category is the same as Chapman. Therefore, NFH dataset can be utilized to evaluate the models trained on the Chapman dataset, verifying the generalization across databases. Moreover, NFH dataset is clinical database originating from hospital. It can provide a relatively accurate reflection of the algorithm's performance in real-world clinical practice. \par

\begin{table*}[htbp]\rmfamily
	\centering
	\caption{Categories and details of utilized databases. The categories are Sinus Bradycardia (SB), Sinus Rhythm (SR), Atrial Fibrillation (AFIB), Grouped Supraventricular Tachycardia (GSVT), Norm (N), Myocardial Infarction (MI), ST/T Change (STTC), Conduction Disturbance (CD), Hypertrophy (HYP), Premature Ventricular Contractions(PVC), Premature Atrial Contractions (PAC), Tachycardia (TACH), and Bradycardia (BRAD).}
	\begin{tabular}{cccccc}
		\toprule
		Database & Class & Categories & Training Set & Test Set  & Validation Set \\
		\midrule
		PTB-XL & 5     & N, MI, STTC, CD, HYP & 12978 & 1642  & 1652 \\
		Chapman & 4     & SB, SR, AFIB, GSVT & 8176  & 1029  & 1022 \\
		SNPH & 5     & N, PVC, PAC, TACH, BRAD & 7592 & 949  & 950 \\
		NFH & 4     & SB, SR, AFIB, GSVT & 27924 & 3490  & 3491 \\
		\bottomrule
	\end{tabular}%
	\label{tab:addlabel3}%
\end{table*}%

Specifically, the data pre-processing stage mainly contains 3 steps: data resampling, data filtering, and data normalization. The databases are all composed of 12-lead ECGs with a 500Hz sampling rate. With the prerequisite of retaining enough details for feature extraction, the ECG signals are down-sampled to 250Hz. Consequently, the training cost is reduced, and the processed data is of acceptable quality for analysis simultaneously. Concerning noise and artifact filtering, a wavelet decomposition method proposed in \cite{martis2013ecg} is utilized. In detail, ECGs are decomposed into approximate components ($A$) and detail components ($D$) by 9-level db6 wavelet decomposition. Then useless components are removed, which are $A_9$, $D_1$, and $D_2$ in accordance with \cite{martis2013ecg}. The rationale behind removing these components is as follows: components $D_1$ and $D_2$ represent the frequency band of 50-60 Hz. Within this range, two of three main ECG noise and artifacts, power-line interference and electromyographic artifacts, usually exist. For the rest one that baseline wondering, it is with 0.5-0.6 Hz frequency. Such range belongs to the component $A_9$, that is the reason why $A_9$ is removed. After that, the remained components are synthesized to reconstruct signals, yielding the filtered ECG signals. Finally, considering the differences in individual variations, measurement devices, environmental factors, etc., the filtered signal will undergo Z-score normalization to unify the data scale. The Z-score is formulated as Eq. (5), where x, z, $\mu$, and $\sigma$ denote the input data, output data, average, and standard deviation, respectively. \par

\begin{flalign}
	&&
	z=\frac{x-\mu}{\sigma}
	&&
\end{flalign}

\subsection{Evaluation Metrics}
To comprehensively evaluate the models from different perspectives, 5 evaluation metrics are utilized in the experiments, including accuracy ($Acc$), precision ($Pre$), recall $rec$, $F_1$-score ($F_1$), and area under the receiver operating characteristic curve ($AUC$). In detail, $Acc$, $Pre$, $Rec0$, and $F_1$ can be formulated from Eq. (6) to (9), where TP, TN, FP, and FN denote true positive samples, true negative samples, false positive samples, and false negative samples, respectively. \par
\begin{flalign}
	&&
	Acc=\frac{TP+TN}{TP+TN+FP+FN}
	&&
\end{flalign}
\vspace{-3ex}
\begin{flalign}
	&&
	F_1=\frac{2\cdot{Pre}\cdot{Rec}}{Pre+Rec}
	&&
\end{flalign}
\vspace{-3ex}
\begin{flalign}
	&&
	Pre=\frac{TP}{TP+FP}
	&&
\end{flalign}
\vspace{-3ex}
\begin{flalign}
	&&
	Rec=\frac{TP}{TP+FN}
	&&
\end{flalign}
$AUC$ is related to the receiver operating characteristic curve (ROC). The vertical and horizontal coordinates of $ROC$ are TP and TN, and $AUC$ is the area under ROC. Therefore, $AUC$ is a more comprehensive metric that can reflect the relationship of sensitivity and specificity. Considering the situation of data imbalance, $Acc$ is more easy to be affected. Thus, $Acc$ could be regarded as a direct reflection of performances. From another perspective, $F_1$ and $AUC$ are more comprehensive and rigorous. To some degree, as the results of $Acc$, $F_1$ and $AUC$ conflict, prioritize the results of $F_1$ and $AUC$.

\subsection{Ablations}
Since the most appropriate architectures in different clinical scenarios might vary, the hyper-parameters that control the model composition are worth evaluating. In these ablation experiments, the segment length, scanning steps, feature dimensions, Mamba block depth, and the bi-directional mechanism are evaluated.
\subsubsection{Ablations on Segment Length and Scanning Step}
Among the hyper-parameters, the segment length might be the most important one. The rationale is that the segment length depends on the receptive fields. For morphologically and rhythmic abnormal CVDs, the best receptive fields might be varied, because what should be focused on could be local features or global features. In addition to segment length, the scanning step is evaluated simultaneously. The reason is that its setting is related to the segment length in this experiment. Specifically, the scanning steps are not set specific values, but the multiple of segment length, including 1, 1/2, and 1/4 times. Moreover, the scanning step is also significant as it determines the interaction between two segments. It's a trade-off between model efficiency and accuracy because shorter steps lead to more information and computations. The experimental results are shown in Fig. (5), the evaluation metric is $F_1$-score. In this experiment, the segment length $p$ is set as 25, 50, 100, and 200, respectively. The length of the whole sequence is 2500. $s$ denotes the scanning step. Note that hyper-parameters are evaluated one after another, thus depth and dimension are set as default values of 12 and 192.\par

\begin{figure*}[htbp]
	\centering	
	\includegraphics[width=\textwidth]{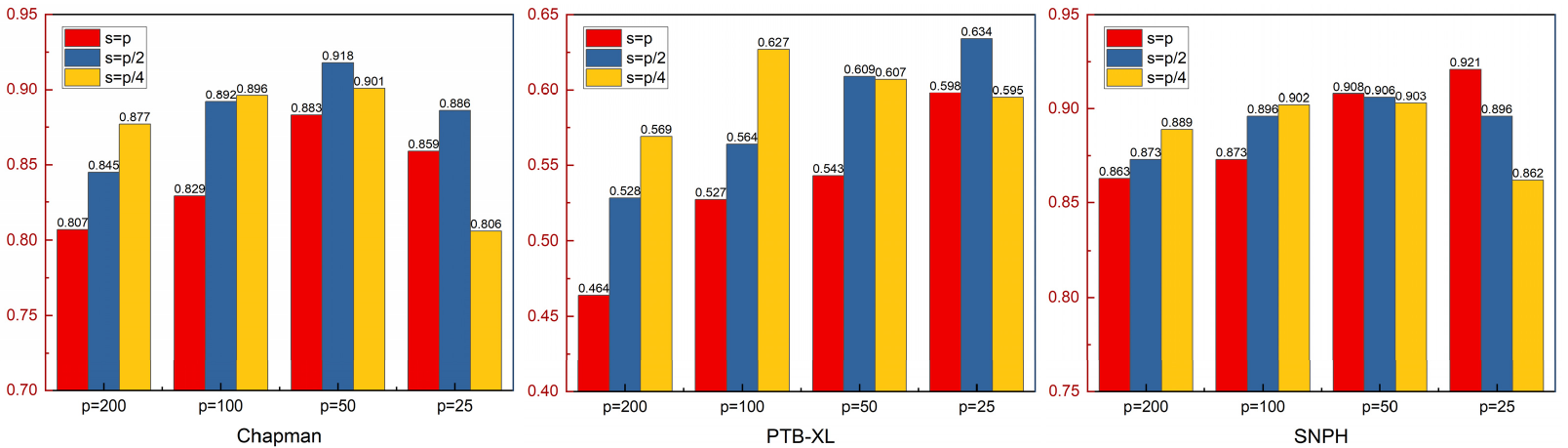}
	\caption{Ablation results on segment length and scanning step.}\label{Figure 5}
\end{figure*}

According to the experimental results, a segment length of 50 sample points is the most appropriate for the Chapman database, while the number for the PTB-XL database is 25. The reasons are that CVDs in Chapman have rhythmic features, thereby some global information matters more, like R-R interval. Diagnosing morphological CVDs in PTB-XL needs more local information, such as P-wave disappearance. Smaller segment length helps the model to concentrate on delicate details, while the longer one is beneficial for extracting global temporal features. Another finding is that performances of S\textsuperscript{2}M\textsuperscript{2}ECG with different segment lengths are closer in the SNPH database. The rationale behind this is its CVDs with mixed categories. \par 
Concerning the analysis scanning step, it should not be evaluated individually. Segment length should be analyzed simultaneously since the value of the scanning step is bound to it. The first phenomenon is a trade-off between segment length and the scanning step. As the segment length is too large, smaller scanning steps are needed, so that the model can guarantee enough focus on delicately local characteristics. As the segment length becomes smaller, the balance point starts to move. For instance, a segment length of 200 usually needs a scanning step that $p/4$, while a segment length of 25 needs $p/2$ or even $p$ scanning steps. In other words, the best balance of the two hyper-parameters yields the best performance. Such balance varies among databases, for Chapman is ($p=50, s=p/2$) , for PTB-XL is ($p=25, s=p/2$), while for SNPH is ($p=25, s=p$).

\subsubsection{Ablations on Mamba Block Depth}
Another important parameter is model depth, it determine the model complexity. Usually, deeper networks could extract higher dimensional features, but they are prone to over-fitting. Considering the performance and cost, moderate model depth is better for algorithm deployment and should be evaluated. In this experiment, the model depth is equal to the number of Mamba blocks, which is set as 6, 12, 18, and 24, respectively. The experimental results are shown in Table 3. \par
\begin{table*}[htbp]
	\centering
	\caption{Experimental results on Mamba block depth.}
	\begin{tabular}{ccccccc}
		\toprule
		Database & Mamba Block Depth & \textit{Accuracy} & \textit{$F_1$-score} & \textit{AUC} & \textit{Precision} & \textit{Recall} \\
		\midrule
		\multirow{4}[2]{*}{Chapman} & 6     & 0.896  & 0.904  & \textbf{0.987}  & 0.904  & 0.905  \\
		& 12    & \textbf{0.928} & \textbf{0.918} & 0.981 & \textbf{0.922} & \textbf{0.922} \\
		& 18    & 0.909  & 0.917  & 0.984  & 0.921  & 0.913  \\
		& 24    & 0.860  & 0.866  & 0.973  & 0.859  & 0.874  \\
		\midrule
		\multirow{4}[2]{*}{PTB-XL} & 6     & \textbf{0.750} & 0.625  & 0.881  & 0.673  & 0.602  \\
		& 12    & 0.740  & \textbf{0.634} & \textbf{0.894} & \textbf{0.659} & \textbf{0.632} \\
		& 18    & 0.742  & 0.631  & 0.876  & 0.673  & 0.604  \\
		& 24    & 0.706  & 0.578  & 0.877  & 0.613  & 0.572  \\
		\midrule
		\multirow{4}[2]{*}{SNPH} & 6     & 0.902  & 0.889  & 0.976  & 0.887  & 0.892  \\
		& 12    & \textbf{0.913} & \textbf{0.921} & \textbf{0.985} & \textbf{0.921} & \textbf{0.916} \\
		& 18    & 0.909  & 0.895  & 0.972  & 0.896  & 0.893  \\
		& 24    & 0.891  & 0.875  & 0.973  & 0.874  & 0.876  \\
		\bottomrule
	\end{tabular}%
	\label{tab:addlabel4}%
\end{table*}%
The results indicate that 12 Mamba blocks nearly outperform all the other situations. Therefore for most cases, the parameter is better to be set as 12. Another phenomenon is that fewer Mamba blocks are enough to unleash the potential of SSMs in feature extraction. Deeper networks (depth=18 or 24) achieve inferior performance instead. The reason is that too high dimensional features are limited to be beneficial for CVD diagnosis in most cases, they are even easily trigger over-fitting. Furthermore, it's also the limitation of Transformer-based models in CVD diagnosis. SSM-based models are only with linear computing complexity, while Transformer-based models are with quadratic ones. In other words, such complexity is too much for CVD diagnosis by ECGs. The simplification from Transformers to SSMs is helpful for model performance and efficient computing. \par

\subsubsection{Ablations on Feature Dimensions}
The feature dimension is another significant parameter that determines the variation of features. Thus, it's evaluated in this experiment, with feature dimensions being set as 24, 48, 96, 192, and 384. The experimental results are presented in Fig. 6, where the evaluation metrics are accuracy ($Acc$), $F_1$-score, area under the curve (AUC), precision ($Pre$), and recall ($Rec$). 

\begin{figure*}[htbp]
	\centering	
	\includegraphics[width=\textwidth]{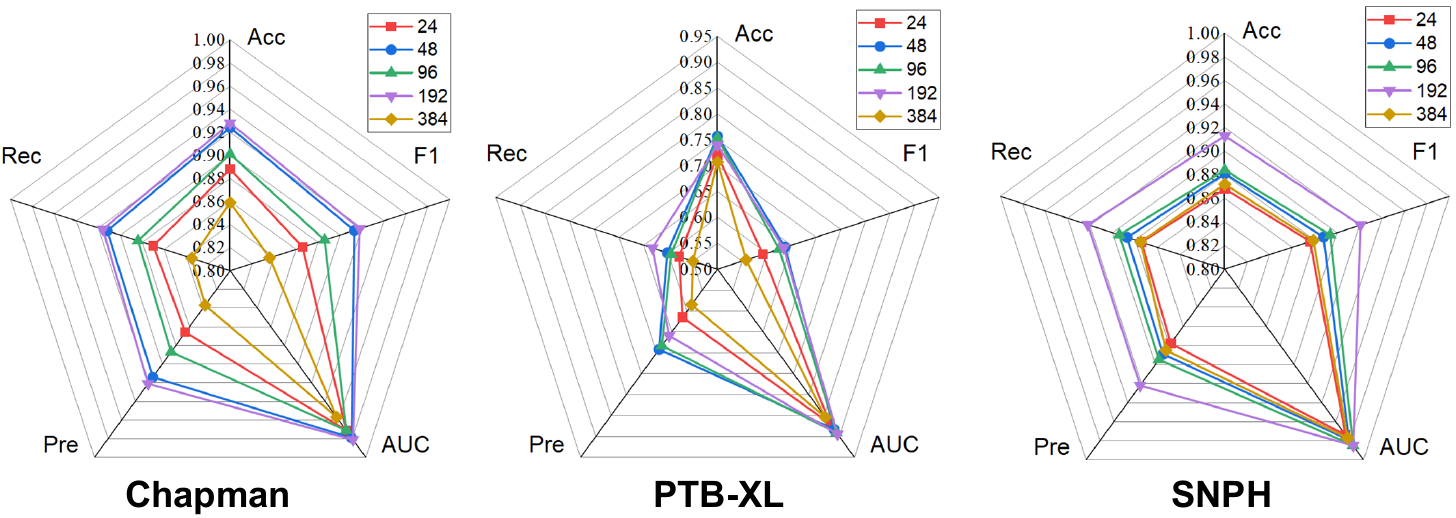}
	\caption{Ablation results on feature dimensions.}\label{Figure 6}
\end{figure*}

The results indicate that the best feature dimensions differentiate from databases. For the Chapman and the SNPH dataset, dimension 192 outperforms other cases, while dimension 48 is the best choice in the PTB-XL scenario. The reasons are related to feature representations. Since the features represent the whole layer, they are based on global feature aggregation. Thus, for the dataset that global features matter, more parameters are needed to represent such global information. From another perspective, for datasets like PTB-XL, local information is important as the feature being aggregated within a window, rather than being incorporated in the global situation.

\subsubsection{Ablations on Bi-directional Scanning}
Last but not least, the designation of bi-directional scanning should be evaluated. Similar to a decoder-only Transformer like GPT, the scanning in vanilla Mamba is single direction. It means that a token could only see its prior tokens, rather than the later ones. For language, such designation works because it represents the semantic analyzing process. However, in the domain of ECGs, information that is hidden in the prior and later curves are both matters. They are utilized for comprehensive analysis together. Thus, the scanning direction should be bi-directional, making a token could simultaneously see its prior and later tokens. In this experiment, such bi-directional scanning is evaluated, as shown in Fig. 7. Note that the other parameters are set the same as the best performance model in particular datasets. \par 

\begin{figure*}[htbp]
	\centering	
	\includegraphics[width=\textwidth]{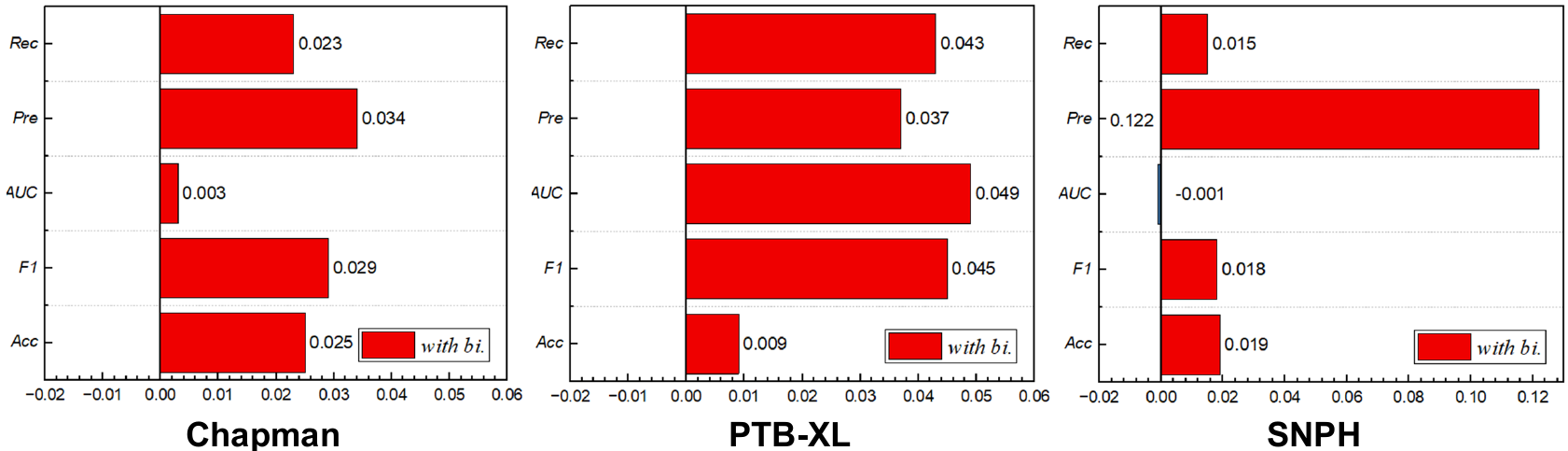}
	\caption{Ablation results on bi-directional scanning designation.}\label{Figure 7}
\end{figure*}

It could be found that bi-directional scanning improves the model performance nearly on all the evaluation metrics. Such enhancement is related to correlation among features and visibility of features. \par 
First of all, there is a corroborative relationship between the prior and later features. Similar to bi-LSTM, bi-directional SSMs provide forward and backward features that are significant for temporal information extraction. Because some valuable information is hidden in both the forward and reverse direction. For instance, before the symptoms of acute myocardial infarction appear, the ECG waveform may exhibit characteristic dynamic evolution, such as ST-segment elevation, formation of pathological Q waves\cite{delewi2013pathological}, etc. Thereby, in forward scanning, these precursors are associated with the symptoms that appear in the subsequent waveform. In reverse, for example, before the onset of dilated cardiomyopathy, some non-specific ST-T abnormalities or changes in the voltage of the QRS complex may occur\cite{silvetti2023pivotal}. Therefore, in reverse scanning, the symptoms of the disease are associated with the arrhythmia that appeared in the previous waveform. It could be found that in the process of scanning, the immediately visible existing knowledge affects the feature extraction of invisible waves. Considering that the visible and invisible parts are determined on the scanning direction, both the forward and reverse ones are crucial. Moreover, some temporal features could only be completely composed based on both the forward and backward features, such as prolonged Q-T interval and elevated S-T segment.\par 
From another perspective, the visibility of features is related to the masking mechanisms of Mamba. Similar to auto-regressive Transformers, Mamba processes tokens one by one with the help of the mask during scanning. The processed tokens are regarded as existing knowledge. Thereby, the order of processing tokens is important, which is decided by the scanning directions. In other words, different scanning directions provide diverse inferences based on various existing knowledge. Reverse scanning becomes a very important supplement for the comprehensive construction of features. Finally, such designation yields obvious enhancements.

\subsection{Roles of ECG-specific Multi-branch Designation}
In this work, all the leads are independent before the lead fusion module. Each lead has its individual encoders and SSM blocks. The rationale behind this is the ECG-specific multi-branch architecture. Owing to the ECG multi-lead system, ECG data can be formatted as aforementioned sub-2D data. Independently analyzing each lead can excavate its particular features without disturbing by other leads. For instance, limb leads are related to the electrical differences among limbs, their features might be quite different from the chest ones. In contrast, ECG data can also be formatted as 1D data with 12 channels, which point to 12 leads. Thus, in this experiment, S\textsuperscript{2}M\textsuperscript{2}ECG with/without multi-branch architectures, i.e., sub-2D and 1D designations, are compared. The experimental results are shown in Table 4.

\begin{table*}[htbp]
	\centering
	\caption{Comparative experiments on S\textsuperscript{2}M\textsuperscript{2}ECG w/o ECG-specific multi-branch designation.}
	\begin{tabular}{ccccccc}
		\toprule
		Database & w/o multi-branch & \textit{Accuracy} & \textit{F1}-score & \textit{AUROC} & \textit{Precision} & \textit{Recall} \\
		\midrule
		\multirow{2}[2]{*}{Chapman} & without & 0.877  & 0.891  & 0.970  & 0.879  & 0.895  \\
		& with  & \textbf{0.913} & \textbf{0.918} & \textbf{0.985} & \textbf{0.922} & \textbf{0.922} \\
		\midrule
		\multirow{2}[2]{*}{PTB-XL} & without & 0.727  & 0.593  & 0.865  & 0.582  & 0.591  \\
		& with  & \textbf{0.757} & \textbf{0.637} & \textbf{0.883} & \textbf{0.692} & \textbf{0.602} \\
		\midrule
		\multirow{2}[2]{*}{SNPH} & without & 0.883  & 0.850  & 0.970  & 0.869  & 0.877  \\
		& with  & \textbf{0.928} & \textbf{0.921} & \textbf{0.981} & \textbf{0.921} & \textbf{0.916} \\
		\bottomrule
	\end{tabular}%
	\label{tab:addlabel5}%
\end{table*}%
According to the experimental results, it could be seen that S\textsuperscript{2}M\textsuperscript{2}ECG without multi-branch architectures achieves a comprehensive increase in performance. Such designation provides independent analyzing process for different leads, so that lead-specific characteristics are more easily to be extracted. After that, all these features are sent to a information fusion module for synthesis. Therefore, lead-specific characteristics are better excavated in independently analyzing process, and global features are extracted during the fusion of features from various leads. In other words, incorporating ECG-specific multi-branch mechanisms in SSM-based models can introduce notable performance improvements.

\subsection{Roles of Lead Fusion Module}
The lead fusion module is another crucial part of S\textsuperscript{2}M\textsuperscript{2}ECG because it provides spatio-temporal information fusion to effectively mingle individual features originating from Mamba blocks. Therefore, in this experiment, S\textsuperscript{2}M\textsuperscript{2}ECG with lead fusion module is compared with that merely with simple concatenation. The experimental results are shown in Table 5. \par

\begin{table*}[htbp]
	\centering
	\caption{Experimental results on the roles of Lead Fusion Module.}
	\begin{tabular}{ccccccc}
		\toprule
		Database & w/o Lead Fusion Module & \textit{Accuracy} & \textit{$F_1$-score} & \textit{AUROC} & \textit{Precision} & \textit{Recall} \\
		\midrule
		\multirow{2}[2]{*}{Chapman} & without & 0.982  & 0.895  & 0.975  & 0.899  & 0.878  \\
		& with  & \textbf{0.913} & \textbf{0.918} & \textbf{0.985} & \textbf{0.922} & \textbf{0.922} \\
		\midrule
		\multirow{2}[2]{*}{PTB-XL} & without & 0.731  & 0.603  & 0.870  & 0.599  & 0.601  \\
		& with  & \textbf{0.757} & \textbf{0.637} & \textbf{0.883} & \textbf{0.692} & \textbf{0.602} \\
		\midrule
		\multirow{2}[2]{*}{SNPH} & without & 0.890  & 0.866  & 0.971  & 0.872  & 0.865  \\
		& with  & \textbf{0.928} & \textbf{0.921} & \textbf{0.981} & \textbf{0.921} & \textbf{0.916} \\
		\bottomrule
	\end{tabular}%
	\label{tab:addlabel6}%
\end{table*}%

The results indicate that the lead fusion module brings an all-around enhancement, no matter in the morphological datasets or the rhythmic ones. The rationale behind this is the spatio-temporal information fusion. The FFN part of the lead fusion module effectively enriches the feature diversity and provides nonlinear transformations on a temporal scale. In this case, temporal features are well aggregated, leading to desirable representations of temporal features. On the other hand, the SENet part makes the cross-lead fusion to be channel-wise. Features from different leads are well-analyzed and weighted. The significance of various inputs dynamically changes to better represent the more important leads to particular CVD cases. Therefore, the lead fusion module and the multi-branch designation are two modules that mutually promote each other and can bring out the greatest potential.\par

\subsection{Clinical generalization}
To evaluate S\textsuperscript{2}M\textsuperscript{2}ECG's general capabilities across clinical databases, a clinical generalization experiment is introduced. Considering that S\textsuperscript{2}M\textsuperscript{2}ECG is a supervised learning method, experiments on generalization across datasets can only be conducted as the utilized datasets are with the same categories. Therefore, the Chapman and NFH databases are employed for the experiments, since they have the exactly same categories. To verify the generalization, S\textsuperscript{2}M\textsuperscript{2}ECG is trained and validated in a database, and then transferred to another for test. The experimental results are shown in Table 6.\par 

\begin{table}[htbp]
	\centering
	\caption{Generalization experiments on Chapman and NFH databases. Chap. and Val. are shorts for Chapman and validation.}
	\adjustbox{width=0.5\textwidth}{
	\begin{tabular}{ccccccc}
		\toprule
		Train \& Val. & Test  & \textit{Acc} & \textit{$F_1$} & \textit{AUC} & \textit{Pre} & \textit{Rec} \\
		\midrule
		Chap. & Chap. & 0.913  & 0.918  & 0.985  & 0.922  & 0.922  \\
		NFH   & Chap. & 0.917  & 0.922  & 0.989  & 0.925  & 0.923  \\
		\midrule
		NFH   & NFH   & 0.940  & 0.926  & 0.985  & 0.925  & 0.929  \\
		Chap. & NFH   & 0.935  & 0.917  & 0.973  & 0.920  & 0.915  \\
		\bottomrule
	\end{tabular}%
	}
	\label{tab:addlabel7}%
\end{table}%

The results reveal the generalization capability of S\textsuperscript{2}M\textsuperscript{2}ECG across clinical datasets. The model maintains satisfactory performance with minimal degradation during migrations, whether from Chapman to NFH or from the NFH database to Chapman. Note that the performances are even maintain an increase as S\textsuperscript{2}M\textsuperscript{2}ECG transferred from NFH to the Chapman database. It stems from the NFH database's larger scale and richer data, enabling the model to improve its analysis of these categories. Given its strong performance in generalization tests on clinical datasets, the model is considered to have robust generalization capabilities and holds potential for deployment in clinical practice or edge devices. \par

\subsection{Comparative Experiments}
As mentioned before, several existing automatic CVD diagnosis methods are based on different deep-learning structures. It's valuable to directly compare S\textsuperscript{2}M\textsuperscript{2}ECG with such methods. Thus, in this experiment, comparative experiments are conducted to evaluate the performance of various models. Specifically, the existing work can be categorized as follows: CNN\cite{he2016deep, hannun2019cardiologist, avetisyan2024deep}, Transformer\cite{vaswani2017attention}, GNN\cite{zhang2023st}, and hybrid models\cite{yao2020multi,he2019automatic, chen2022automated, xie2022multilabel, ji2024msgformer}. The experimental results are shown in Table 7.

\begin{table*}[htbp]
	\centering
	\caption{Results of the comparative experiments. Higher $Acc$, $F_1$, and $AUC$ denote better performances, lower Model Size (parameters) means lighter model weight.}
	\adjustbox{width=\textwidth}{
		\renewcommand{\arraystretch}{1.25}
		\begin{tabular}{ccccccccccccccl}
			\toprule
			\multirow{2}[4]{*}{Method} & \multirow{2}[4]{*}{Structure} & \multicolumn{3}{c}{Chapman} &       & \multicolumn{3}{c}{PTB-XL} &       & \multicolumn{3}{c}{SNPH} &       & \multicolumn{1}{c}{\multirow{2}[4]{*}{\makecell{Model Size$\downarrow$\\(parameters)}}} \\
			\cmidrule{3-5}\cmidrule{7-9}\cmidrule{11-13}          &       & \textit{Acc} & $F_1$   & \textit{AUC} &       & \textit{Acc} & $F_1$    & \textit{AUC} &       & \textit{Acc} & $F_1$   & \textit{AUC} &       &  \\
			\midrule
			ResNet-18\cite{he2016deep} & \multirow{3}[2]{*}{CNN} & 0.784  & 0.741  & 0.924  &       & 0.767  & 0.644  & 0.867  &       & 0.905  & 0.914  & \textbf{0.988} &       & 3.854M \\
			\cite{hannun2019cardiologist} &       & 0.889  & 0.872  & 0.970  &       & 0.772  & 0.639  & 0.843  &       & 0.904  & 0.908  & 0.983  &       & 39.193M \\
			\cite{avetisyan2024deep} &       & 0.872  & 0.853  & 0.967  &       & \textbf{0.782} & \textbf{0.649} & \textbf{0.884} &       & 0.884  & 0.890  & 0.891  &       & 3.004M \\
			\midrule
			\cite{yao2020multi} & \multirow{4}[2]{*}{CNN+RNN} & 0.699  & 0.648  & 0.844  &       & 0.773  & 0.629  & 0.818  &       & 0.900  & 0.899  & 0.978  &       & 1.779M \\
			\cite{he2019automatic} &       & 0.891  & 0.874  & 0.970  &       & 0.777  & 0.635  & 0.837  &       & 0.906  & 0.911  & 0.984  &       & 0.948M \\
			\cite{chen2022automated} &       & 0.854  & 0.827  & 0.958  &       & 0.760  & 0.621  & 0.860  &       & 0.895  & 0.901  & 0.950  &       & 6.702M \\
			\cite{xie2022multilabel} &       & 0.779  & 0.736  & 0.920  &       & 0.668  & 0.509  & 0.785  &       & 0.832  & 0.844  & 0.867  &       & 6.440M \\
			\midrule
			ViT\cite{vaswani2017attention}   & Trans. & 0.738  & 0.694  & 0.887  &       & 0.609  & 0.525  & 0.808  &       & 0.865  & 0.860  & 0.874  &       & 12.751M \\
			MSGformer\cite{ji2024msgformer} & CNN+Trans. & 0.726  & 0.672  & 0.871  &       & 0.687  & 0.519  & 0.777  &       & 0.863  & 0.855  & 0.871  &       & 19.052M \\
			ST-ReGE\cite{zhang2023st} & GNN   & 0.886  & 0.863  & 0.971  &       & 0.756  & 0.624  & 0.876  &       & 0.900  & 0.903  & 0.979  &       & \textbf{0.097M} \\
			S\textsuperscript{2}M\textsuperscript{2}ECG & SSM   & \textbf{0.928} & \textbf{0.918} & \textbf{0.981} &       & 0.757  & 0.637  & 0.883  &       & \textbf{0.913} & \textbf{0.921} & 0.985  &       & 0.705M \\
			\bottomrule
		\end{tabular}%
		\label{tab:addlabel8}%
	}
\end{table*}%

According to the results, the proposed S\textsuperscript{2}M\textsuperscript{2}ECG shows favorable performances in all cases. As evaluated on the Chapman database, S\textsuperscript{2}M\textsuperscript{2}ECG outperforms existing methods at least by 4.15\% in \textit{Acc}, 5.03\% in $F_1$, 1.03\% in \textit{AUC}. The advantages become 0.77\% in \textit{Acc}, 0.77\% in $F_1$ when it comes to the SNPH database. And the difference between S\textsuperscript{2}M\textsuperscript{2}ECG and the best-performing models is only 0.003. With respect to the PTB-XL database, S\textsuperscript{2}M\textsuperscript{2}ECG also achieves a performance close to that of the best-performing model. The performance gap between them is not significant, merely 0.025 in \textit{Acc}, 0.012 in $F_1$, and 0.001 in \textit{AUC}. Meanwhile, the number of parameters of S\textsuperscript{2}M\textsuperscript{2}ECG is only about 1/30 of that of the best-performing model. \par
Considering that the performances of existing methods are significantly related to their structures, it's worth comparing S\textsuperscript{2}M\textsuperscript{2}ECG with models based on each structure. First, as mentioned before, CNN-based models are good at capturing local ECG morphological features through their translation-invariant convolutional filters. This is also reflected in the experimental results. CNN-based models\cite{he2016deep, hannun2019cardiologist, avetisyan2024deep} show superior performances in PTB-XL, the morphological CVD datasets. While S\textsuperscript{2}M\textsuperscript{2}ECG's state-space modeling prioritizes long-range rhythm analysis, resulting in better performance in rhythmic ones. From the parameter aspect, S\textsuperscript{2}M\textsuperscript{2}ECG achieves at least a 4.26× parameter reduction over CNN-based models attributed to its state-sharing mechanism and input-dependent selective computation. Secondly, comparing to the RNN-based or hybrid ones\cite{yao2020multi, he2019automatic, chen2022automated, xie2022multilabel}, S\textsuperscript{2}M\textsuperscript{2}ECG achieves an obvious outperforming. The reason is that S\textsuperscript{2}M\textsuperscript{2}ECG has a similar ability to model long-range spatio-temporal dependencies without RNNs' vanishing gradient limitations and inefficient serial computation. And comparing to the ones based on Transformers\cite{vaswani2017attention, ji2024msgformer}, S\textsuperscript{2}M\textsuperscript{2}ECG has linear-complexity state-space modeling that simultaneously captures long-range temporal rhythms and multi-lead spatial correlations without attention's quadratic bottlenecks. Consequently, S\textsuperscript{2}M\textsuperscript{2}ECG achieves distinct superiority with at least an 18.09× parameter reduction. Finally, as compared to GNN-based model\cite{zhang2023st}, S\textsuperscript{2}M\textsuperscript{2}ECG achieves superior ECG spatio-temporal modeling by dynamically learning lead correlations through state-space transitions, eliminating GNNs' reliance on predefined graph structures that cannot adapt to patient-specific ECG patterns. While GNNs require fewer parameters by sharing graph convolution kernels across all leads, this compactness comes at the cost of rigid topological assumptions that degrade morphological sensitivity. As mentioned before, the differences in parameters are acceptable because such GNN-based methods are designed for extremely lightweight models with receptive performances. To sum up, S\textsuperscript{2}M\textsuperscript{2}ECG simultaneously achieves long-term dependency, parallel computing, linear computational complexity, and dynamic lead correlations. These properties make S\textsuperscript{2}M\textsuperscript{2}ECG a competitive alternative in many scenarios.\par

It could be found that S\textsuperscript{2}M\textsuperscript{2}ECG does not achieve an all-around enhancement in all the databases. It does not outperform all the other existing models in the morphological dataset, i.e., PTB-XL. Although the gap between S\textsuperscript{2}M\textsuperscript{2}ECG and the best-performing model is not significant, which is 0.012 in $F_1$ and 0.001 in \textit{AUC}, and the current performance is acceptable, the reasons for this under-performing are still worthy of discussion. Essentially, the reasons are more related to intrinsic characteristics of temporal feature extraction among different architectures. Three fundamental factors may contribute to the sub-optimality of S\textsuperscript{2}M\textsuperscript{2}ECG in morphological tasks. Firstly, local pattern sensitivity. ECG morphology interpretation relies heavily on localized waveform characteristics. However, S\textsuperscript{2}M\textsuperscript{2}ECG's global state space modeling might overlook subtle localized morphological variations. In reverse, for some models like CNN-based ones, their inductive bias through convolutional kernels enables automatic learning of translation-invariant local features across hierarchical levels. This will help the waveform feature extraction. Secondly, multi-scale feature integration. Morphological abnormalities often manifest through coordinated changes at different temporal scales. But S\textsuperscript{2}M\textsuperscript{2}ECG's uniform scaling mechanism may insufficiently capture the complex interplay between short-term waveform details and longer morphological trends. As for CNN-based models, they inherently combine multi-resolution features through pooling operations and deep layer stacking. It's a natural multi-scale designation. Last but not least, parameter efficiency trade-off. While Mamba demonstrates remarkable parameter efficiency through its selective memory mechanism, this design potentially limits its capacity to model the high-dimensional feature interactions required for precise morphology classification. CNN-based models, despite requiring more parameters, might better leverage ECG's quasi-periodic nature through weight-sharing convolutions.  Thus, S\textsuperscript{2}M\textsuperscript{2}ECG gets better performance on the temporal CVD diagnosis than the morphological ones. But in comprehensive and practical scenarios SNPH, S\textsuperscript{2}M\textsuperscript{2}ECG is more generic. \par

Another important perspective is the parameter volumes. Since increasing computational efficiency and balancing the trade-off of complexity-performance are substantially important features of SSMs, S\textsuperscript{2}M\textsuperscript{2}ECG's advantages in lightweight computing should be emphasized. From the table, it could be seen that S\textsuperscript{2}M\textsuperscript{2}ECG obtains nearly the minimum number of parameters. As mentioned before, this is largely attributed to its ability to process time-series data with linear complexity. For one hand, S\textsuperscript{2}M\textsuperscript{2}ECG is the most lightweight designation comparing to those based on CNNs, RNNs, and Transformers, etc. From another hand, lightweight is not equal to decrease in performance. Considering the similarity of Transformers and SSMs, ViT and MSGformer, for instance, are more appropriate for comparisons. To some degree, the most significant difference between them is that the self-attention mechanism in Transformers has exponential complexity for tokens, while SSM only has linear one. However, higher complexity does not bring superior performances. S\textsuperscript{2}M\textsuperscript{2}ECG outperforms ViT and MSGformer in all databases, and meanwhile, it merely obtains about 10\% parameters of them. This result is consistent with the previous discussion on "moderate" complexity. It can be said that S\textsuperscript{2}M\textsuperscript{2}ECG's linear complexity is more suitable for sub-2D data like ECG, striking a balance between high performance and lightweight. It's worth mentioning that ST-ReGE is a graph-based model that pursues extremely lightweight and acceptable performance. However, from the performance views, S\textsuperscript{2}M\textsuperscript{2}ECG acquires obvious advantages over ST-ReGE in all the databases. To sum up, S\textsuperscript{2}M\textsuperscript{2}ECG achieves superior performance in temporal scenarios and comparable performance in morphological situations, and it also gets fewer parameters than nearly all the other existing methods, which is meaningful for algorithm deployments and practices.

\subsection{Statistical significance analysis}
In addition to the average performances, statistical significance tests need to be introduced to increase the confidence of experimental results. First, the results of comparable experiments are given a confidence interval. The formula of the confidence interval is formulated as Eq. (10), where CI is the confidence interval, $\overline{x}$ is the average performance, $t_{\alpha/2}$ is the distribution threshold, s denotes the standard deviation, and $n$ denotes degrees of freedom which is equal to the number of experiments. \par
\begin{flalign}
	&&
    CI=\overline{x}\substack{+ \\ -}t_{\alpha/2}\frac{s}{\sqrt{n}}
	&&
	\end{flalign}\par
In this experiment, A confidence level of 95\% is chosen, thereby $t_{\alpha/2}$ is equal to 2.776, and the number of experiments is 5. Considering that $F_1$-score is a more comprehensive metric to reflect model performances, S\textsuperscript{2}M\textsuperscript{2}ECG is compared with existing methods in $F_1$-score with confidence interval. The results are shown in Table 8. The results demonstrate that S\textsuperscript{2}M\textsuperscript{2}ECG has good stability, small performance fluctuations, and high experimental repeatability. \par

\begin{table}[htbp]
	\centering
	\caption{Comparative experiments with confidence interval of $F_1$-score.}
	\adjustbox{width=0.5\textwidth}{
	\begin{tabular}{cccc}
		\toprule
		Method & Chapman & PTB-XL & SNPH \\
		\midrule
		ResNet-18\cite{he2016deep} & 0.741+0.006 & 0.644+0.005 & 0.914+0.003 \\
		\cite{hannun2019cardiologist} & 0.872+0.005 & 0.639+0.005 & 0.908+0.003 \\
		\cite{avetisyan2024deep} & 0.853+0.004 & \textbf{0.649+0.007} & 0.890+0.001 \\
		\midrule
		\cite{yao2020multi} & 0.648+0.005 & 0.629+0.005 & 0.899+0.005 \\
		\cite{he2019automatic} & 0.874+0.008 & 0.635+0.008 & 0.911+0.005 \\
		\cite{chen2022automated} & 0.827+0.010 & 0.621+0.008 & 0.901+0.008 \\
		\cite{xie2022multilabel} & 0.736+0.005 & 0.509+0.003 & 0.844+0.004 \\
		\midrule
		ViT\cite{vaswani2017attention}   & 0.694+0.004 & 0.525+0.002 & 0.860+0.003 \\
		MSGformer\cite{ji2024msgformer} & 0.672+0.007 & 0.519+0.004 & 0.855+0.004 \\
		ST-ReGE\cite{zhang2023st} & 0.863+0.005 & 0.624+0.009 & 0.903+0.002 \\
		S\textsuperscript{2}M\textsuperscript{2}ECG & \textbf{0.918+0.005} & 0.637+0.005 & \textbf{0.921+0.003} \\
		\bottomrule
	\end{tabular}%
	}
	\label{tab:addlabel1}%
\end{table}%

In addition, since the average performances of S\textsuperscript{2}M\textsuperscript{2}ECG are better than other existing methods in Chapman and SNPH databases, significance testing should be included. Independent samples t-test is employed in this experiment. Similarly, $F_1$-score is chosen as the evaluation metric. S\textsuperscript{2}M\textsuperscript{2}ECG is compared with the best-performing models on the Chapman and SNPH databases, which are methods in \cite{he2019automatic} and \cite{he2016deep}. In the Chapman database, the p-value for significance testing is 3.60$\times10^{-5}$, and that for the SNPH database is 9.09$\times10^{-5}$. Both of the p values are less than the commonly used significance level of 0.05. Therefore, the null hypothesis can be rejected and model performance is determined to be significantly different.\par

\subsection{Experiment of Inference Latency}
According to previous experiments, it could be seen that S\textsuperscript{2}M\textsuperscript{2}ECG is near with fewer parameters, this illustrates its spatial efficiency. Concerning time efficiency, it's worth evaluating S\textsuperscript{2}M\textsuperscript{2}ECG on the inference latency. At the same time, it could also reveal the deploy-ability on edge devices of S\textsuperscript{2}M\textsuperscript{2}ECG. The experiment object is S\textsuperscript{2}M\textsuperscript{2}ECG with different hyper-parameters. According to previous experiments, consistency is achieved in the values of Mamba block depth and feature dimensions for the best-performing model across databases, which are set as 12 and 192. Thus, this experiment is based on S\textsuperscript{2}M\textsuperscript{2}ECG with various patch sizes and scanning steps. \par
First, S\textsuperscript{2}M\textsuperscript{2}ECG with different hyperparameters are transferred into the ONNX file first. It's due to some characteristics of ONNX files, including cross-platform compatibility, model conversion and deployment simplicity, fast inference speed, etc. Another important reason is that the ONNX file is a more generalized form for edge deployments. Then, the transferred models are deployed on the device for inference, which is Intel(R) Xeon(R) CPU E5-2687W in this experiment. The data utilized for inference is a 10-second ECG signal with 250Hz. The experimental results are illustrated in Fig. 8.\par

\begin{figure}[htbp]
	\centering	
	\includegraphics[width=0.45\textwidth]{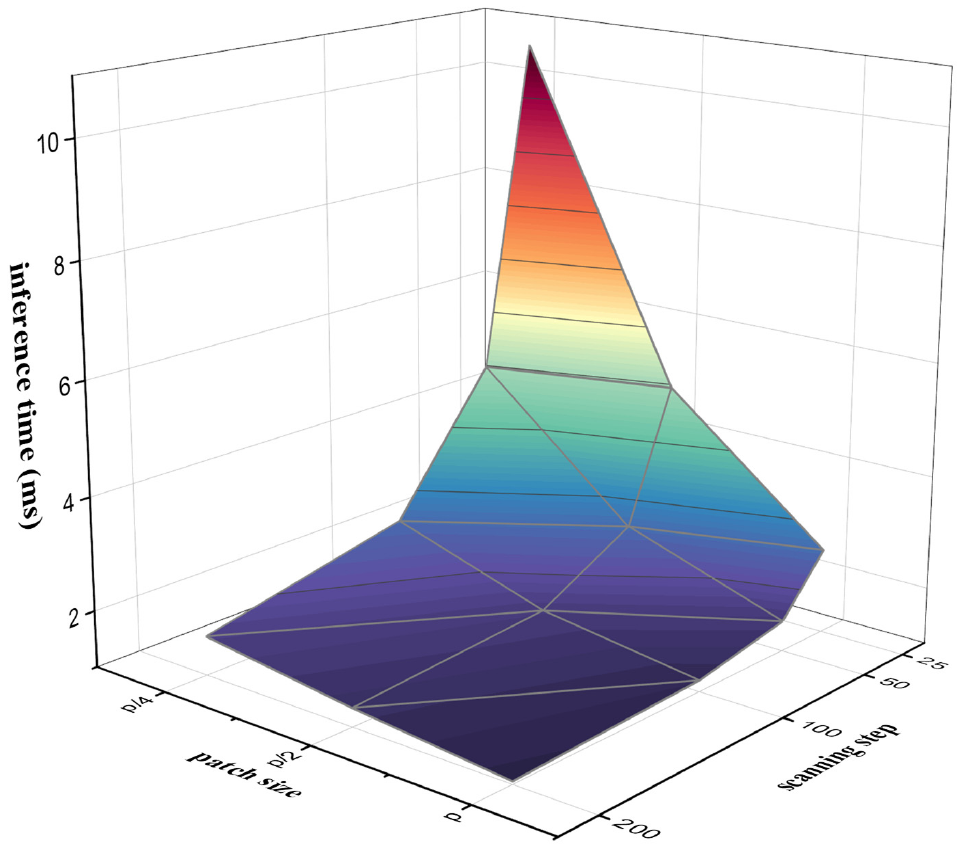}
	\caption{Experiments on inference latency of S\textsuperscript{2}M\textsuperscript{2}ECG.}\label{Figure 8}
\end{figure}
In accordance with the results, it could be found that S\textsuperscript{2}M\textsuperscript{2}ECG with a larger patch size and larger scanning step has higher inference speed. This is understandable as it represents less computation. Another finding is that nearly all of the inference time is less than 5ms, except for the one with the least patch size and the scanning step. However, even the latency of the most time-consuming model fully satisfies the real-time rate requirement. According to \cite{ravi2016deep}, when analyzing a 10-second length ECG signal, if the computation cost obtained is significantly less than the 10-second segments used, the inference is considered to meet real-time requirements. In this experiment, the inference time is from 1.049 to 10.571 milliseconds (ms), which is significantly less than 10 seconds. Additionally, the best-performing models on three databases only cost 2.484ms, 2.495ms, and 4.782ms, respectively. Therefore, the proposed S\textsuperscript{2}M\textsuperscript{2}ECG has the potential to be successfully deployed on edge devices with low latency. \par
In addition to time feasibility, spatial deployability is also worth evaluating. In this experiment, the spatial deployability is evaluated from the memory utilization aspect. The results are shown in Fig. (9). According to the results, the memory utilization is between 1063MB and 1245MB. Since most edge devices can meet such memory requirements, S\textsuperscript{2}M\textsuperscript{2}ECG also has the potential to be successfully deployed on edge devices with low memory utilization.
\begin{figure}[htbp]
	\centering	
	\includegraphics[width=0.45\textwidth]{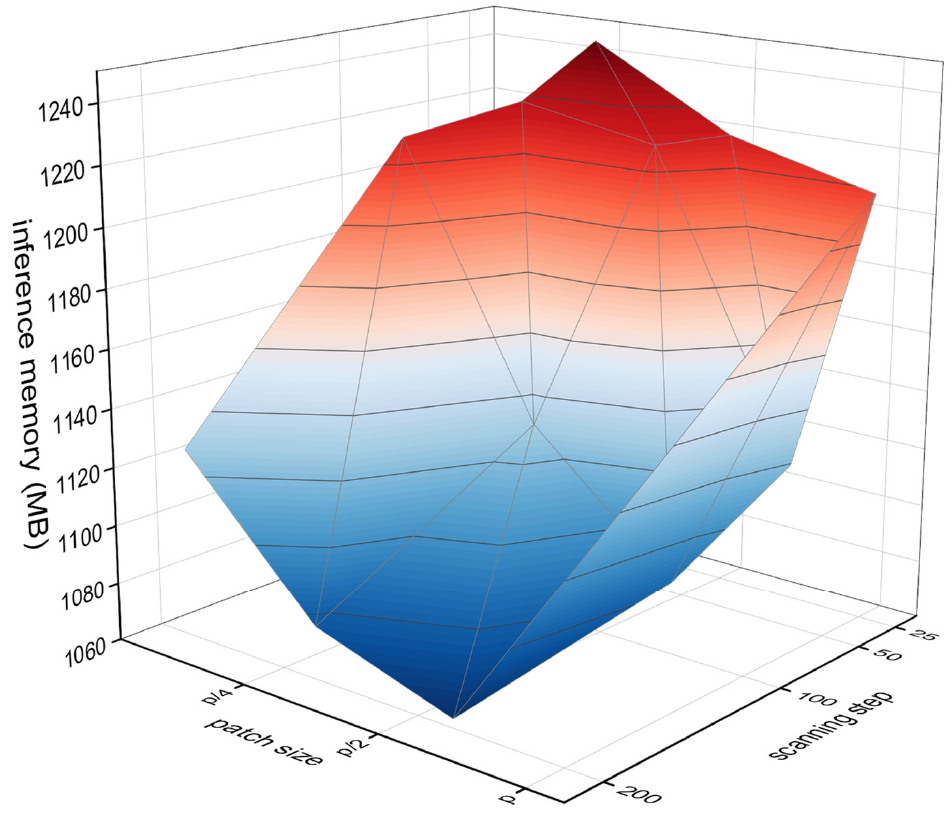}
	\caption{Experiments on inference memory of S\textsuperscript{2}M\textsuperscript{2}ECG.}\label{Figure 9}
\end{figure}

\subsection{Clinical Implications of S\textsuperscript{2}M\textsuperscript{2}ECG}
The superior performance of S\textsuperscript{2}M\textsuperscript{2}ECG on clinical ECG database SNPH carries significant implications for real-world medical applications. Automatic detection in clinical practice may witness some improvements. \par
First, the large-scale clinical ECG analysis in hospitals or diagnostic centers. S\textsuperscript{2}M\textsuperscript{2}ECG’s linear-time complexity enables efficient processing of high-volume ECG records (e.g., 24-hour Holter monitoring with ~100,000 beats). The reduction of the computational costs benefits the effective analysis of such long-term data. S\textsuperscript{2}M\textsuperscript{2}ECG's long-range rhythm modeling brings the potential to improve the detection of intermittent arrhythmia which is often missed in short-duration ECG clips. In addition, considering the scalability of Mamba, in the scenario of automatic detection by large medical devices, S\textsuperscript{2}M\textsuperscript{2}ECG is expected to obtain models with a larger number of parameters and stronger representational capabilities for more accurate CVD diagnosis. Second, for portable and wearable ECG monitoring. With only 0.705M parameters, S\textsuperscript{2}M\textsuperscript{2}ECG can run on low-power edge devices. This brings the potential for convenient ECG monitoring in daily use. The parameter sharing mechanism, selective computation, and hardware-aware optimization of Mamba enable real-time analysis in the portable devices. Finally, S\textsuperscript{2}M\textsuperscript{2}ECG has potential in longitudinal ECG interpretation. S\textsuperscript{2}M\textsuperscript{2}ECG’s continuous-time modeling better captures slow-evolving pathologies. It is beneficial for enhanced long-term trend analysis. \par

\subsection{Limitations and Future Work}
While S\textsuperscript{2}M\textsuperscript{2}ECG demonstrates superior performance in rhythm analysis, clinical analysis, and computational efficiency, there still remain some limitations. S\textsuperscript{2}M\textsuperscript{2}ECG does not achieve an all-around enhancement over other existing work, especially in morphological scenarios. This is attributed to Mamba’s global state-space prioritization, which may overlook fine-grained spatial patterns that CNNs capture via localized convolutional filters. Another limitation is the explainability. While attention maps in Transformers or activation heatmaps in CNNs provide intuitive explanations, Mamba’s state transition dynamics are less visually interpretable, posing challenges for clinical adoption. \par
Beyond algorithmic limitations, this study also acknowledges several data-related constraints. While the SNPH and NFH database employed in this research originates from real-world clinical practice and provides partial validation of S\textsuperscript{2}M\textsuperscript{2}ECG's diagnostic capabilities, it should be noted that actual clinical scenarios are substantially more complex. Clinical practice will include broader demographic diversity, heterogeneity in acquisition devices and data sources, and more granular disease classifications. Consequently, the current simulation cannot comprehensively reflect the model's performance in true clinical settings. Faithful clinical replication would require substantially larger and more heterogeneous datasets.\par 
According to the current limitations, there are some extended works worthy of research in the future. To enhance morphological sensitivity while preserving efficiency, S\textsuperscript{2}M\textsuperscript{2}ECG can be integrated with other architectures like CNNs or Transformers. Such hybrid designation holds the promise of making up for each architecture's shortcomings. In addition, state-space visualization interfaces that highlight critical time steps and lead contributions can be designed, akin to attention mechanisms but tailored for SSMs. Such designation might help enhance the explainability of S\textsuperscript{2}M\textsuperscript{2}ECG. Finally, to better approximate real-world clinical practice, future studies should incorporate more comprehensive training data. First, demographic expansion should be included. Broader age spectrum representation, balanced ethnic/racial distributions, and sex-specific subgroup analyses should be included. Secondly, data from multiple sources should be employed. The data will be from heterogeneous acquisition devices and different hospitals. Thus, multicenter cross-validation protocols could be implemented. At last, more fine-grained data should be utilized for evaluations. And the number of categories will be expanded at the same time.\par

\section{Conclusion}
This study presents an efficient cardiovascular disease (CVD) diagnosis methodology that capitalizes on spatio-temporal bi-directional structured state space models (SSMs) and a multi-branch Mamba architecture. The proposed method effectively combines the advantages of convolutional neural networks (CNNs), Transformers, and the ECG-specific multi-branch feature extraction. \par 
The S\textsuperscript{2}M\textsuperscript{2}ECG architecture is meticulously designed with scanning windows and segment tokenization to aggregate local information and extract morphological features analogous to convolutional kernels. This enables fine-grained analysis of ECG signals. Bidirectional scanning and sequence processing components enable long-range dependency modeling comparable to Transformers, ensuring robust temporal feature extraction essential for rhythm analysis. The multi-branch design uncovers inter-lead relationships through dedicated encoders and fusion modules, offering comprehensive physiological insights. \par 
A key advantage lies in its computational efficiency: S\textsuperscript{2}M\textsuperscript{2}ECG achieves these capabilities with linear complexity, enabling real-time inference and efficient deployment in clinical settings. Extensive evaluations confirm its superior performance in rhythm-based CVD diagnosis, competitive accuracy in morphological analysis, and strong clinical applicability—making it both technically robust and practically viable.\par 
The lightweight parameter count (an order of magnitude smaller than attention-based models) facilitates edge device integration, aligning with the Internet of Medical Things (IoMT) vision for pervasive health monitoring. This hardware compatibility supports real-world implementation on resource-constrained platforms. \par 
In conclusion, S\textsuperscript{2}M\textsuperscript{2}ECG significantly advances by delivering high-performance CVD diagnosis with unprecedented efficiency. The architecture's unique combination of computational lightness, long-range modeling, and multi-lead analysis positions it as a promising solution for scalable, real-time cardiovascular health monitoring.

\section*{Acknowledgments}
This was was supported in part by......

\bibliographystyle{unsrt}  
\bibliography{ref}

\end{document}